\documentclass[11pt]{article}

\usepackage[margin=1.0in]{geometry}
\usepackage{amsmath,amssymb,amsfonts,graphicx}
\usepackage{hyperref}
\usepackage{slashed}
\usepackage{wasysym}
\usepackage{hhline}
\usepackage{tabularx}
\usepackage{utfsym}

\usepackage[utf8]{inputenc}

\usepackage[greek,english]{babel}

\usepackage[caption=false]{subfig}
\usepackage{booktabs}
\usepackage{soul}

\usepackage[font=small,labelfont=bf]{caption}
\usepackage{cite}
\usepackage{enumerate}

\usepackage{xcolor}
\usepackage{bm}

\definecolor{gray}{cmyk}{0,0,0,0.05}

\newcommand{\CM}{\textsc{CheckMATE}}
\newcommand{\spey}{\textsc{Spey}}
\newcommand{\ifb}{\ensuremath{{\text{fb}}^{-1}}}
\newcommand{\met}{\ensuremath{E_{\mathrm{T}}^\mathrm{miss}}}
\newcommand{\CLs}{\ensuremath{\mathrm{CL}_s}}
\newcommand{\T}{\ensuremath{\mathrm{T}}}
\newcommand{\neut}[1]{\ensuremath{\tilde{\chi}_{#1}^0}}

\usepackage{soul}

\begin{document}

\begin{center}
\vskip 1cm
{\LARGE \bf 
Implementation of full and simplified likelihoods in CheckMATE}
\vskip 1.5cm
{\large 
I\~naki Lara and Krzysztof Rolbiecki 
}
\vskip 1cm
{\em
Faculty of Physics,
University of Warsaw\\ ul.~Pasteura 5, PL-02-093 Warsaw, Poland 
}
\vskip 0.3cm
{\texttt{Inaki.Lara@fuw.edu.pl}, \texttt{Krzysztof.Rolbiecki@fuw.edu.pl}}
\end{center}

\vskip 1.5cm

\today
\begin{abstract}
We present the implementation of simplified and full likelihood models for multibin signal regions in \CM{}. A total of 13 searches are included from ATLAS and CMS, and several methods are presented for the implementation and evaluation of likelihood functions. Statistical combinations increase the sensitivity of searches and open up the possibility of combining orthogonal search channels in the \CM{} framework.  
\end{abstract}

\newpage

\section{Introduction\label{sec:1}}

The Run 3 at the Large Hadron Collider (LHC) is now in full swing, and by the end of the current run it is expected that the LHC would collect several hundreds of inverse femtobarns of data. Apart from the Higgs boson discovery\cite{ATLAS:2012yve,CMS:2012qbp}, which made the experimental picture of the Standard Model complete, no sign of new physics has been observed. However, the vast data has only been interpreted in the limited number of theoretical models by the LHC experiments. On many occasions, the interpretation is provided in terms of simplified models~\cite{LHCNewPhysicsWorkingGroup:2011mji} which do not directly correspond to realistic TeV-scale physics. This situation has prompted several groups~\cite{Kraml:2013mwa,Alguero:2021dig,Conte:2018vmg,Bierlich:2019rhm,GAMBIT:2017qxg,Unel:2021edl} to provide computer programs that would allow reinterpretation of experimental results in terms of an arbitrary new physics model~\cite{LHCReinterpretationForum:2020xtr}. 

Over the years, searches for new physics were becoming increasingly sophisticated, including the full use of complicated statistical models~\cite{Cranmer:2021urp} and machine learning methods~\cite{Albertsson:2018maf,Araz:2023mda}. Instead of reporting results in a handful of bins, often optimized for a small class of models, and comparing observed and expected numbers of events, the experiments nowadays provide binned data in several observables in signal and control regions. On the one hand, this can significantly improve the sensitivity, but on the other hand, it greatly complicates the reinterpretation of the results. In this approach, experiments in high-energy physics test the compatibility of collision data with theoretical predictions, which can be described as a likelihood function. The function gives the probability of the data assuming a theoretical model and a certain set of parameters. The likelihood can therefore be used to constrain the parameters of the model.  

In this paper, we discuss the implementation of the full and simplified likelihoods in the \CM{} framework~\cite{Drees:2013wra,Kim:2015wza,Dercks:2016npn,Desai:2021jsa}. For implementations in other tools, see, e.g.~\cite{Alguero:2020grj,Alguero:2022gwm}. In this paper, we include 9 ATLAS and 4 CMS searches. The CMS searches use the correlated background model (covariance matrix)~\cite{Collaboration:2242860}. For ATLAS searches, we use the simplified~\cite{ATL-PHYS-PUB-2021-038} and full likelihood~\cite{ATL-PHYS-PUB-2019-029} frameworks. In addition, some of the implementations of the ATLAS searches include the full set of control regions. Users have several ways of controlling the actual calculation of likelihoods. The default method of evaluation is based on the \spey{} package~\cite{Araz:2023bwx}. We provide validation plots for each of the searches. A comparison of three methods for the calculation of limits is presented: the best-expected signal region, the full likelihood model (if available), and the simplified likelihood.    

The paper is organized as follows. In Section~\ref{sec:technical}, we briefly discuss statistical models and their technical implementation in \CM{}. We provide a list of implemented searches and options available to a user to control the program's behavior. In Section~\ref{sec:validation} we present validation plots for each of the searches discussed. A brief summary and outlook are provided in the last section. 

\section{Technical implementation\label{sec:technical}}

In this section, we introduce the methods for the implementation of simplified and full likelihoods in \CM{} and user switches to control their execution.

\subsection{ATLAS}

The functionality of combining signal regions for recasting in ATLAS searches can be implemented using either the full likelihood model~\cite{ATL-PHYS-PUB-2019-029} or following a simplified approach detailed in Ref.~\cite{ATL-PHYS-PUB-2021-038}. Table~\ref{tab:analysis_atlas} lists the ATLAS analyses with likelihood functionality implemented in \CM{}.  The simplified likelihood method requires background rates and uncertainties that were already available in the implemented searches. The full likelihood requires an appropriate file in the JSON format and these files were released by ATLAS for 6 searches already implemented in \CM{}. For the searches \texttt{atlas\_2004\_14060}, \texttt{atlas\_2006\_05880}, and \texttt{atlas\_2111\_08372} the full model files are not available but using the published data one can still perform a simplified model fitting in multibin signal regions. 

The full-likelihood statistical models are encoded in the JSON files by the ATLAS Collaboration. The files are not shipped with \CM{} but are automatically downloaded during installation from the \href{https://www.hepdata.net/}{\textsc{HEPData}} website~\cite{Buckley:2010jn,Maguire:2017ypu}, \href{https://www.hepdata.net/}{\texttt{https://www.hepdata.net/}}. The information provided includes the number of background events for all signal and control regions and for each major background category separately. This results in a large number of nuisance parameters and the complexity of the procedure makes the hypothesis testing very CPU-expensive. Additionally, on the recasting side, in order to fully exploit the method, one should also implement control regions (CR), which was not a standard approach in \CM{}. Currently, only two searches \texttt{atlas\_2010\_14293} and \texttt{atlas\_1911\_06660} have complete implementation of all CRs. In other searches, it is assumed that the contribution of signal to CRs is negligible. This assumption is not obviously fulfilled in all imaginable new physics models.   

On the technical side, after the usual evaluation of events within \CM{}, a JSON patchset is created that encapsulates signal contributions to signal regions (SRs) (and CRs if applicable). The patchset is then combined with the background-only model from ATLAS. Since the signal region names in \CM{} do not usually exactly match the names in ATLAS conventions, an additional file is required, \texttt{pyhf\_conf.json}, which provides a dictionary between two conventions.  The likelihood is further evaluated using the \textsc{Pyhf}~\cite{Cowan:2010js,pyhf_joss,pyhf} package, which is a Python implementation of the \textsc{HistFactory} specification for binned statistical models~\cite{Cranmer:1456844,Baak:2014wma}. The signal strength $\mu$ is the parameter of interest, where $\mu = 1$ corresponds to the nominal cross section of a tested model. Depending on the user's choices, the output can contain information about the expected and observed upper limit on $\mu$, with 2-$\sigma$ bounds, along with the observed and expected $\mathrm{CL}_s$ for $\mu=1$, where
\begin{equation}
    \CLs = \frac{\mathrm{CL}_{s+b}}{\mathrm{CL}_b}
\end{equation}
and $\mathrm{CL}_{s+b}$, $\mathrm{CL}_b$  are the $p$-values of the signal and the null hypotheses, respectively.
Asymptotic formulae are the default calculation method; see~\cite{Cowan:2010js}. 

By default, the above calculation will be executed using the \spey{} program~\cite{Araz:2023bwx} and the \textsc{Spey-Pyhf} plugin~\cite{speypyhf}. \spey{} is a Python-based cross-platform package that allows for statistical inference of hypotheses using different likelihood prescriptions.\footnote{Installation of \spey{} is straightforward: \texttt{pip install spey}. Please refer to \spey{} \href{https://speysidehep.github.io/spey/quick_start.html}{online documentation} and \textsc{Spey-Pyhf} for more details~\cite{speyonline}.} In our setup, it gives a somewhat better control over the calculation than the above-mentioned \CM-\textsc{Pyhf} interface, but nevertheless the calculation is still performed in the \textsc{Pyhf} framework. However, the main motivation for using \spey{} was the possibility of combining different searches (also between experiments), which is planned in the next release of \CM. In any case, a direct evaluation using \textsc{Pyhf} and bypassing \spey{} is also available.   

Since the evaluation of full likelihoods is normally time-consuming, it is not practical for large scans of the parameter space. Therefore, the alternative approach is based on the concept of simplified likelihood~\cite{ATL-PHYS-PUB-2021-038}. In this case, the background model is approximated with the total SM background rate obtained in the background-only fit in the full model. A single nuisance parameter correlated over all bins and representing post-fit background uncertainty is constrained by the unit normal distribution. The evaluation is also performed using the \textsc{Pyhf} package. As such, a simplified likelihood can usually be constructed from publicly available information for many ATLAS studies.\footnote{Another approach was adopted in \textsc{MadAnalysis}: the \textsc{Simplify} tool~\cite{simplify} is used to construct simplified likelihoods from full likelihood models, and technically both approaches should be equivalent.} In \CM{} this is performed on the fly in a dedicated routine. This feature should be used with caution, and we advise cross-checking with the full likelihood calculation. In certain situations, the simplified likelihood can result in significant over-constraining of models, as it was shown in Ref.~\cite{Lahiri:2025opz} (Appendix A2) in the case of invisible Higgs boson decays in an implementation of the ATLAS analysis~\cite{ATLAS:2022yvh}. By default, \CM{} performs full likelihood fits and the simplified method has to be activated using \texttt{Model:~simple}, see Tab.~\ref{tab:cmparams}.

\newcolumntype{H}{>{\setbox0=\hbox\bgroup}c<{\egroup}@{}}
\begin{table}
{\small
\begin{tabularx}{\textwidth}
{lXrrcc}
\toprule
Name & Description  &   \#SR & N$_\mathrm{bin}$   &  Full  & Ref.  \\ \midrule
 \texttt{atlas\_1908\_03122} & Search for bottom squarks in final states with Higgs bosons, $b$-jets and \met & 2& 7 
 & \usym{1F5F8}& \cite{ATLAS:2019gdh}\\
 \texttt{atlas\_1908\_08215} & Search for electroweak production of charginos and sleptons in final states with 2 leptons and \met & 1& 52 
 & \usym{1F5F8} & \cite{ATLAS:2019lff}\\
 \texttt{atlas\_1911\_06660} & Search for direct stau production in events with two hadronic taus & 1& 2 
 &  \usym{1F5F8} & \cite{ATLAS:2019gti} \\
 \texttt{atlas\_1911\_12606} & Search for electroweak production of supersymmetric particles with compressed mass spectra & 2& 76 
 & \usym{1F5F8} & \cite{ATLAS:2019lng}\\
 \texttt{atlas\_2004\_14060} & Search for stops in hadronic final states with \met & 3& 14 & \usym{1F5F4} & \cite{ATLAS:2020dsf}   \\
 \texttt{atlas\_2006\_05880} & Search for top squarks in events with a Higgs or $Z$ boson & 3& 23 & \usym{1F5F4} &  \cite{ATLAS:2020aci}   \\
 \texttt{atlas\_2010\_14293} & Search for squarks and gluinos in final states with jets and \met & 3& 60 
 & \usym{1F5F8} & \cite{ATLAS:2020syg} \\
 \texttt{atlas\_2101\_01629} & Search for squarks and gluinos in final states with one isolated lepton, jets, and \met & 1& 26
 & \usym{1F5F8} & \cite{ATLAS:2021twp} \\
 %
 \texttt{atlas\_2111\_08372} & Search for associated production of a $Z$ boson with
an invisibly decaying Higgs boson or dark matter candidates & 1& 22 & \usym{1F5F4} & \cite{ATLAS:2021gcn} \\
\bottomrule
\end{tabularx}
\caption{List of implemented ATLAS analyses which have likelihood-based signal regions (all searches at $\sqrt{s} = 13$~TeV and $\mathcal{L} = 139$ \ifb). 
\label{tab:analysis_atlas}
}
}
\end{table}

\subsection{CMS}

The simplified likelihood framework was defined in Ref.~\cite{Collaboration:2242860}. This assumes correlation between background contributions that can be modeled using the multivariate Gaussian distribution:
\begin{equation}
 \mathcal{L}_S(\mu, \boldsymbol\theta) = \prod_{i=1}^{N} \frac{(\mu \cdot s_i + b_i + \theta_i)^{n_i} e^{-(\mu \cdot s_i + b_i + \theta_i)}}{n_i!} \cdot \exp\left(-\frac{1}{2} \boldsymbol\theta^T \mathbf{V}^{-1} \boldsymbol\theta \right)
\end{equation}
where the product runs over all bins and $\mu$ is the signal strength (and the Parameter of Interest - POI), $n_i$ the observed number of events, $s_i$ the expected number of signal events, $b_i$ the expected number of background events, $\theta_i$ a background nuisance parameter, and $\mathbf{V}$ the covariance matrix. It is implemented using the covariance matrices provided by the CMS Collaboration, which are included in the \CM{} distribution in the JSON format. The evaluation of the above model is performed using the \spey{} package and the \texttt{default\_pdf.correlated\_background} method.

\newcolumntype{H}{>{\setbox0=\hbox\bgroup}c<{\egroup}@{}}
\begin{table}
{\small
\begin{tabularx}{\textwidth}{lXcl}
\toprule
Name & Description  &   N$_\mathrm{bin}$  & Ref.  \\ \midrule
\texttt{cms\_1908\_04722} & Search for supersymmetry in final states with jets and \met & 174 & \cite{CMS:2019zmd} \\
\texttt{cms\_1909\_03460} & Search for supersymmetry with $M_\mathrm{T2}$ variable in final states with jets and \met & 282 & \cite{CMS:2019ybf} \\
\texttt{cms\_2107\_13021} & Search for new particles in events with energetic jets and large \met & 66 & \cite{CMS:2021far}\\
\texttt{cms\_2205\_09597} & Search for production of charginos and neutralinos 
 in final states containing hadronic decays of $WW$, $WZ$, or $WH$ and \met & 35 & \cite{CMS:2021far}\\
\bottomrule
\end{tabularx}
\caption{List of implemented CMS analyses which have likelihood-based signal regions (all searches at $\sqrt{s} = 13$~TeV and $\mathcal{L} = 139$ \ifb). 
\label{tab:analysis_cms}
}
}
\end{table}

\subsection{\CM{} parameters}

\CM{} provides several switches and parameters to control the details of the statistical evaluation. These are summarized in Tab.~\ref{tab:cmparams}. The switches are divided into two groups: one providing a control of what statistical tests are performed, and the other to control different modes of calculation. By default, no statistical evaluation is performed. For the sake of speed and stability, one switch, \texttt{scan}, provides a quick and reliable way of obtaining an \texttt{Allowed/Excluded} result but with limited additional information. Generally, available statistics include \CLs{} tests and calculation of upper limits on signal strength, both of which can be obtained as observed and/or expected measures. By choosing a \texttt{select} switch, users can control which statistics are calculated. If no explicit choice is made, the \textit{observed upper limit} will be calculated. Finally, the \texttt{detailed} switch can be used to request calculation of all available statistics, but it should be noted that its execution can be time consuming. The option \texttt{-so} can be used to request the calculation of statistics for previous \CM{} runs (it requires the presence of the \texttt{evaluation/total\_results.dat} file in the output directory). 

The second group of parameters is used to choose a method of calculation of requested statistics for the ATLAS searches (it does not affect the calculation for the CMS searches as described in the previous Section). For the default method, the \texttt{full} switch chooses the calculation using the full likelihood and the \CM-\textsc{Spey} interface. With the \texttt{simple} switch, the calculation is performed using simplified likelihood and the \CM-\textsc{Pyhf} interface. Finally, the \texttt{fullpyhf} switch requests the calculation using the full likelihood and \CM-\textsc{Pyhf} interface (this is somewhat less flexible with regard to the output compared to the previous options). In any case, users should remember that the full likelihood calculation can be time consuming if many searches and signal regions are requested.   

\begin{table}
\begin{tabularx}{\textwidth}{lllX}
\toprule
Parameter card & Terminal & \texttt{X} & Description and available choices \\
\toprule
\texttt{Multibin: X} &  \texttt{-mb X} & \texttt{none} & No signal region combination is performed (default).  \\
 &   & \texttt{select} & Calculates user selected statistics.   \\
 &   & \texttt{scan} & Calculates observed \CLs{}; fast and reliable for quick assessment of exclusion.   \\
  &   & \texttt{detailed} & Calculates observed and expected upper limits and \CLs{}.  \\ 
\texttt{Expected: False} & \texttt{-exp} & & Selects calculation of expected limits. \\
\texttt{CLs: \hfill False} & \texttt{-mbcls} & & Selects calculation of \CLs{}. \\
\texttt{Uplim: \hfill False} & \texttt{-uplim} & & Selects calculation of upper limits. \\
\texttt{Statonly: False} & \texttt{-so} & & Calculates statistical combinations without event-level analysis provided the analysis and evaluation steps were already completed. \\
\midrule 
& & \texttt{full} & The \spey interface to the full likelihood model for ATLAS searches (default).  \\
\texttt{Model: X} &  \texttt{-mod X} & \texttt{simple} & The simplified likelihood model for ATLAS searches. \\ 
& & \texttt{fullpyhf} & The full likelihood model for ATLAS searches with \textsc{Pyhf} interface.  \\ 
\midrule
\texttt{Backend: X} &  & \texttt{numpy} & Different backends for calculations in the \texttt{full} model; \textsc{Numpy} is always available. \\  
& & \texttt{pytorch} &  \textsc{Pytorch} backend (if available) \\
& & \texttt{tensorflow} &  \textsc{Tensorflow} backend (if available) \\ 
& & \texttt{jax} &  \textsc{Jax} backend (if available) \\ 
\bottomrule
\end{tabularx}
\caption{Summary of options related to multibin signal regions. 
\label{tab:cmparams}}
\end{table}

The results of the calculation for each of the multibin signal regions and all requested analyses are stored in the \texttt{multibin\_limits/results.dat} file. In order to follow the progress of the calculation, the observed limits are also displayed on the screen for each of the signal regions. 
The final evaluation is decided using the upper limit on the signal strength $\mu$ (if calculated):
\begin{eqnarray*}
 && \mu < 1\  \Longrightarrow\ \texttt{\color{red}Excluded}\\
 && \mu \ge 1\  \Longrightarrow\ \texttt{\color{green}Allowed}.
\end{eqnarray*}
If the results for the upper limit are not available, the decision is made using the observed \CLs{} statistics at the 95\% confidence level:
\begin{eqnarray*}
 && \CLs < 0.05\  \Longrightarrow\ \texttt{\color{red}Excluded}\\
 && \CLs \ge 0.05\  \Longrightarrow\ \texttt{\color{green}Allowed}.
\end{eqnarray*}

\subsection{Evaluation time}
Four different backends for the calculation of full likelihood models in the \textsc{pyhf}/\textsc{Spey} setup are supported: \textsc{Numpy}~\cite{numpy}, \textsc{Tensorflow}~\cite{tf}, \textsc{Pytorch}~\cite{pytorch}, and \textsc{Jax}~\cite{jax}. The \textsc{Numpy} backend does not require additional system components to be installed, but it generally is not recommended due to long computation time. When no backend is explicitly selected, \CM{} will check if the \textsc{Jax} backend is available. The \textsc{Jax} backend was also successfully tested with NVIDIA CUDA library and GPU support.

An example calculation was performed for the evaluation of the full likelihood model in the ATLAS search \texttt{atlas\_2101\_01629}, which is a moderately complex one of those implemented. The process considered was $pp\to \tilde{g}\tilde{g} \to qqqqWW\tilde{\chi}^0_1\tilde{\chi}^0_1$ with an intermediate chargino; masses $m_{\tilde{g}} = 2200$~GeV, $m_{\tilde{\chi}^\pm_1} = 1100$~GeV, $m_{\tilde{\chi}^0_1} = 1$~GeV; cross section $\sigma = 0.46$~fb. The sample size was 1000 events though the times in Tab.~\ref{tab:performance} do not include the sample analysis time. The tests were performed on three Ubuntu~22.04.5 systems: A - Intel i7-4770 CPU 3.4 GHz, 4/8 cores/threads;\footnote{It is really old.} B - Intel i7-8565U CPU 1.8 GHz, 4/8 cores/threads and NVIDIA GeForce MX150; C - Intel i7-7700 CPU 3.4 GHz, 4/8 cores/threads. \textsc{Jax}, \textsc{Pytorch}, and \textsc{Tensorflow} take advantage of multithreading even when there are no GPUs present. On system B the CUDA~12.8 toolkit was installed for GPU computation. In Table~\ref{tab:performance} we show the results. Note that system B was tested in two configurations, with and without GPUs. The best performance is obtained with \textsc{Jax} and \textsc{Pytorch}. For comparison, we also provide evaluation times for the \CLs{} observed limit performed in system A. The results agreed between different calculation methods.

As already mentioned, for large scans of parameter space it may not be practical to perform full likelihood evaluation. The simplified models improve the computation and detailed calculation of all signal strength limits and \CLs{} takes in our example 16 seconds for the \textsc{Jax} backend and 64 seconds for the \textsc{Numpy} backend. Finally, if the calculation of \CLs{} is sufficient, it only takes about 1~second. 

\begin{table}\centering
\begin{tabularx}{0.65\textwidth}{l|llll|l}
\toprule
& \multicolumn{4}{>{\hsize=\dimexpr4\hsize+6\tabcolsep}c|}{upper limits observed and expected} &  \CLs \\ \midrule 
Backend & CPU A & CPU B & GPU B & CPU C & CPU A \\ \midrule
\textsc{Numpy} & 11092 & 12621 & N/A & 8879 & 636\\
\textsc{Jax} & 723 & 526 & 353 & 311 & 55\\
\textsc{Pytorch} & 583 & 805 & N/A & 404 & 57\\
\textsc{Tensorflow} & 2426 & 2920 & 1906 & 1808 & 156 \\ \bottomrule
\end{tabularx}
\caption{Performance comparison (time in seconds) of different \textsc{Pyhf} backends for calculation of the full likelihood in the analysis \texttt{atlas\_2101\_01629}. Systems A and C run on (multithread) CPU, while system B was tested on CPU and GPU. As a reference in the last column the computation time of \CLs{} observed limit is shown. \label{tab:performance}}
\end{table}

\section{Validation\label{sec:validation}}

Multibin signal regions are currently available in 9 ATLAS and 4 CMS analyses, as listed in Tabs.~\ref{tab:analysis_atlas} and \ref{tab:analysis_cms}. The searches are based on the full Run 2 luminosity of about 140~\ifb\ at the center-of-mass energy $\sqrt{s} = 13\ \mathrm{TeV}$. In this section, we briefly introduce each of the searches and provide validation examples. When possible, we compare the full and simplified likelihood approaches, and provide examples of how the multibin evaluation improves exclusion limits compared to the best-signal-region approach. The best-SR (BSR) is defined as the signal region with the best \textit{expected} sensitivity.  

The validation procedure is organized as follows. For SUSY processes, events are generated using \textsc{MadGraph5\_aMC@NLO 3.1.0}~\cite{Alwall:2014hca,Alwall:2007fs,Alwall:2008qv} and the \texttt{MSSM\_SLHA2}~\cite{Duhr:2011se} UFO~\cite{Darme:2023jdn} model with up to two additional partons in the final state. The NNPDF23LO~\cite{Ball:2012cx,Buckley:2014ana,NNPDF:2014otw} parton distribution function (PDF) set is used. The events are then interfaced with \textsc{Pythia}~8.3~\cite{Sjostrand:2014zea,Bierlich:2022pfr} to model decays, hadronization, and showering. The matrix element and parton shower matching for ATLAS searches was performed using the CKKW-L~\cite{Lonnblad:2011xx} prescription and a matching scale of 1/4 of the SUSY particle mass was set. For CMS searches, the MLM~\cite{Mangano:2006rw,Alwall:2007fs} matching scheme was used. In the final step, a fast detector simulation is performed using \textsc{Delphes~3.5}~\cite{deFavereau:2013fsa}. Jets are reconstructed using \textsc{FastJet}~\cite{Cacciari:2011ma} and the anti-$k_t$ algorithm~\cite{Cacciari:2008gp}.  

Inclusive signal cross sections for the production of squarks and gluinos are obtained at the approximate next-to-next-to-leading order with soft gluon resummation at the next-to-next-to-leading-logarithm order (approximate NNLO+NNLL)~\cite{Beenakker:2016lwe,Beenakker:1996ch,Kulesza:2008jb,Kulesza:2009kq,Beenakker:2009ha,Beenakker:2011sf,Beenakker:2013mva,Beenakker:2014sma,Beenakker:1997ut,Beenakker:2010nq,Beenakker:2016gmf}, following the recommendations of Ref.~\cite{Butterworth:2015oua}. The signal cross sections for production of sleptons, charginos and neutralinos are computed at next-to-leading order plus next-to-leading-log precision using \textsc{Resummino}~\cite{Fuks:2013vua,Fiaschi:2018xdm,Bozzi:2007tea,Bozzi:2007qr,Fiaschi:2018hgm,Fuks:2012qx,Debove:2011xj}. This setup generally follows procedures employed within LHC experiments to obtain exclusion limits used in the validation process. 

Events for the simplified scenario of DM production in the CMS search EXO-20-004~\cite{CMS:2021far} were generated at NLO with \textsc{MadGraph5\_aMC@NLO 3.1.0} and the \textsc{DMsimp} model~\cite{Backovic:2015soa}. Up to 2 additional partons were included and the FxFx~\cite{Frederix:2012ps} jet matching scheme was used. The samples for the 2HDM+a model~\cite{Bauer:2017ota} and the validation of HIGG-2018-26~\cite{ATLAS:2021gcn} were simulated internally in \CM{} interfaced with \textsc{MadGraph5\_aMC@NLO 3.4.1} and hadronized with \textsc{Pythia}~8.2.

The sampling of parameter space generally follows grids provided by experiments in the supplementary materials on \textsc{HEPData}. The number of Monte-Carlo (MC) events per point varies; however, close to exclusion lines at mass limits, we aim at samples 10-20 times larger than the nominal luminosity derived values. The exclusion lines are then interpolated based on the grids of simulated parameter points, with additional sampling points occasionally added. MC errors are consistently included in the \textsc{pyhf} patch files.   

For the purpose of validation, we present comparisons between \CM{} and the official ATLAS result of both expected and observed exclusion limits. Where available, the results for simplified and full likelihood are provided. In addition, we also include exclusion contours using the best-\textit{expected} signal region. Note that this usually is based on comparing observed signal with the post-fit background-only prediction in a given bin, which typically has reduced uncertainty, and hence can be quite an aggressive approach. Also, a limit obtained with the full likelihood model is not a combination of such bin-by-bin measurements and cannot be compared in a straightforward way. On the other hand, the simplified likelihood will be such a combination in most cases and it is not surprising that it will produce stronger constraints than the full likelihood in some searches; see also Ref.~\cite{Araz:2023bwx}  The latter generally shows weaker sensitivity than multibin histogram shape-fits, but can be advantageous in terms of evaluation speed.   

We follow the same key for the plots throughout this section. The black solid (dashed) line is for the observed (expected) ATLAS/CMS limit. On some occasions, this is accompanied by theoretical (experimental) $\pm 1\sigma$ uncertainty. The red lines are used for the exclusion obtained with \CM{} using the best signal region method, solid and dashed for the observed and expected limits, respectively. Finally, the blue and green lines denote the limits obtained using the simplified and full likelihoods, respectively.  

\subsection{\texttt{atlas\_1908\_03122} (SUSY-2018-31)}
This is a search \cite{ATLAS:2019gdh} for bottom squark production in final states containing Higgs bosons, $b$-jets, and missing transverse momentum. The Higgs boson is reconstructed from two $b$-tagged jets. The final states contain at least 3 (SRC) or 4 (SRA, SRB) $b$-jets, no leptons, and large missing transverse momentum. The signal region A is divided into 3 bins according to effective mass, $m_\textrm{eff}$, and the signal region C is divided into 4 bins of missing transverse energy significance, $\mathcal{S}$. Thus, both SRA and SRC allow for a shape-fit analysis. Three full likelihood models are provided for the signal regions SRA, SRB, and SRC. 

The validation plots for this search are presented in Fig.~\ref{fig:190803122}. We compare bottom squark pair production, $pp\to \tilde{b}_1 \tilde{b}_1^*$, and two distinct mass spectra. The first assumes that the mass of the lightest supersymmetric particle (LSP) is $m_{\neut{1}} = 60$~GeV and the other assumes the mass difference between neutralinos $\Delta m(\neut{2},\neut{1}) = 130$~GeV. Generally a good agreement between \CM{} and ATLAS is observed for the shape-fit method with \CM{} being slightly weaker. In particular, there is a very good agreement between simplified and full models. The best SR method performs well for the model with fixed LSP mass, however, in the second scenario its exclusion strength can be up to 3 times weaker than shape-fits.

\begin{figure}
 \includegraphics[width=0.5\textwidth]{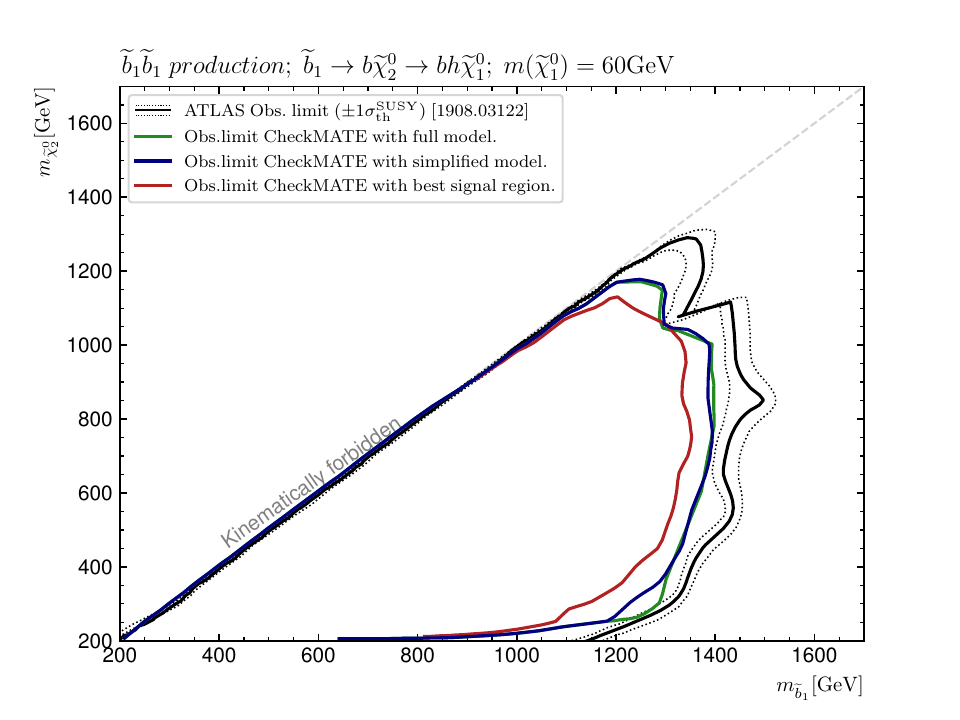}
 \includegraphics[width=0.5\textwidth]{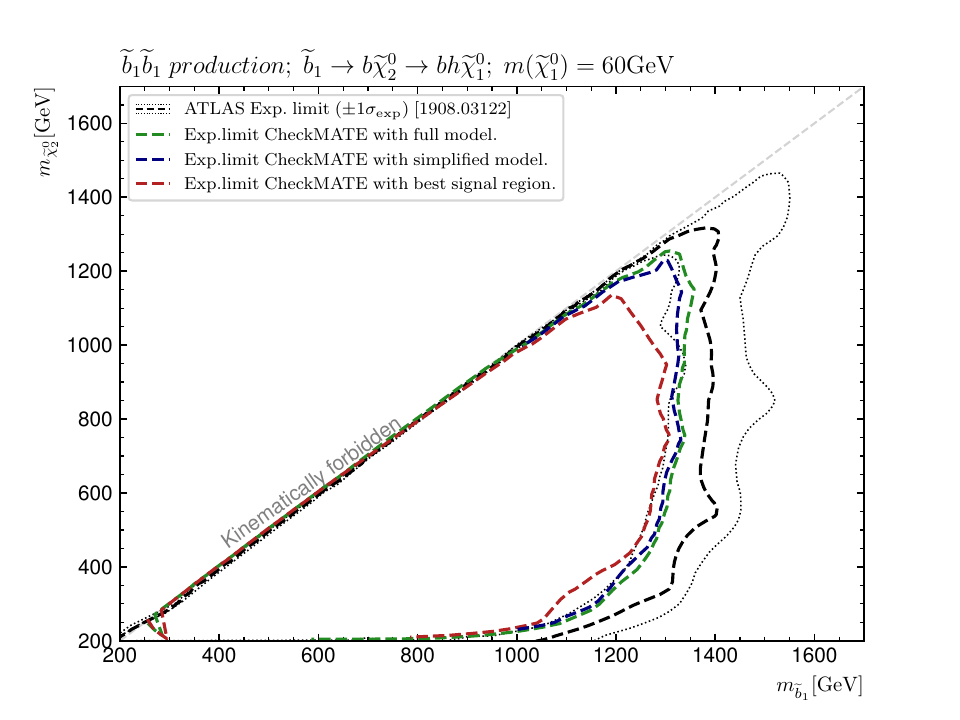}\\
  \includegraphics[width=0.5\textwidth]{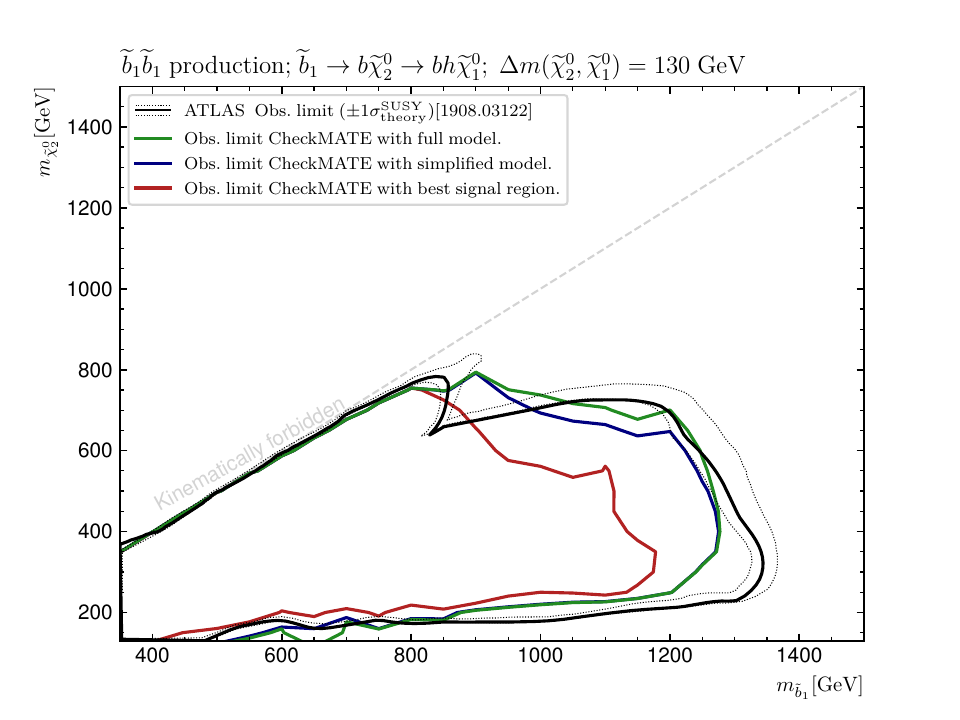}
 \includegraphics[width=0.5\textwidth]{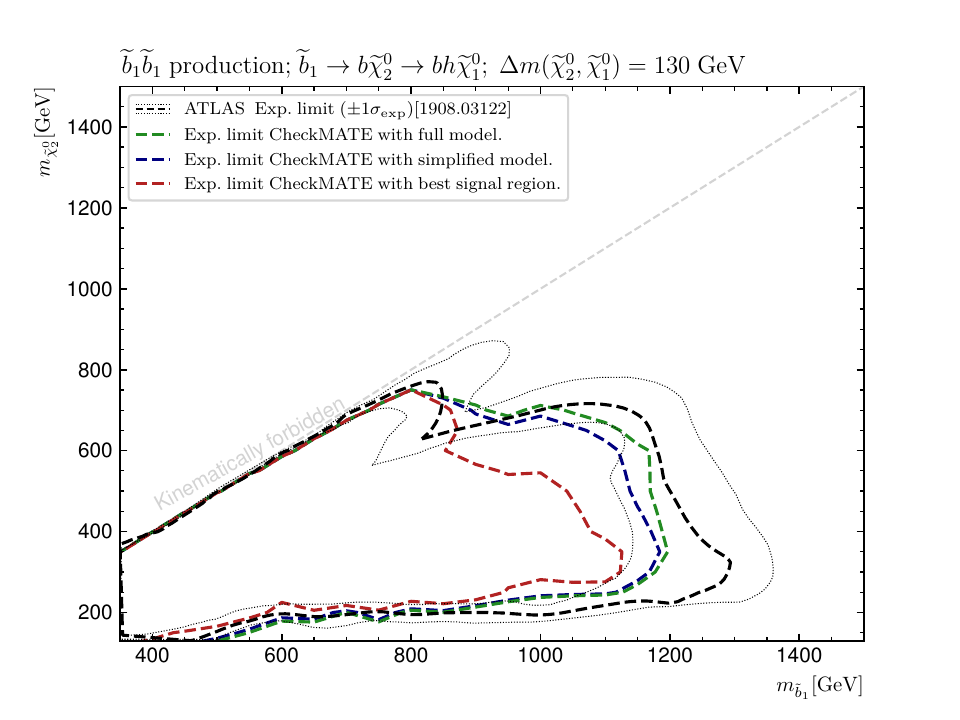}
 \caption{Validation plots for the search \texttt{atlas\_1908\_03122} (SUSY-2018-31). Top row: the model with $m_{\neut{1}} = 60$~GeV; bottom row: the model with $\Delta m(\neut{2},\neut{1}) = 130$~GeV. Left panels: observed limits; right panels: expected limits.\label{fig:190803122}}
\end{figure}

\subsection{\texttt{atlas\_1908\_08215} (SUSY-2018-32)}
This is a search \cite{ATLAS:2019lff} for the production of electroweakinos and sleptons. The final states with two leptons (opposite sign and same or different flavor SF/DF) and \met are considered. Events are first separated into SF and DF categories and are further subdivided by the multiplicity of the non-b-tagged jets. A combined multi-bin SR is defined out of the 36 exclusive binned signal regions. The full likelihood statistical model is provided.

The validation plots for the production of slepton pairs are shown in Fig.~\ref{fig:190808215}. Both the full- and simplified likelihood models agree very well with the ATLAS result. The BSR method gives visibly weaker constraints. Additional validation material for \CM, including cut-flows, can be found in~\cite{190803122_validation}. 

\begin{figure}
\includegraphics[width=0.5\textwidth]{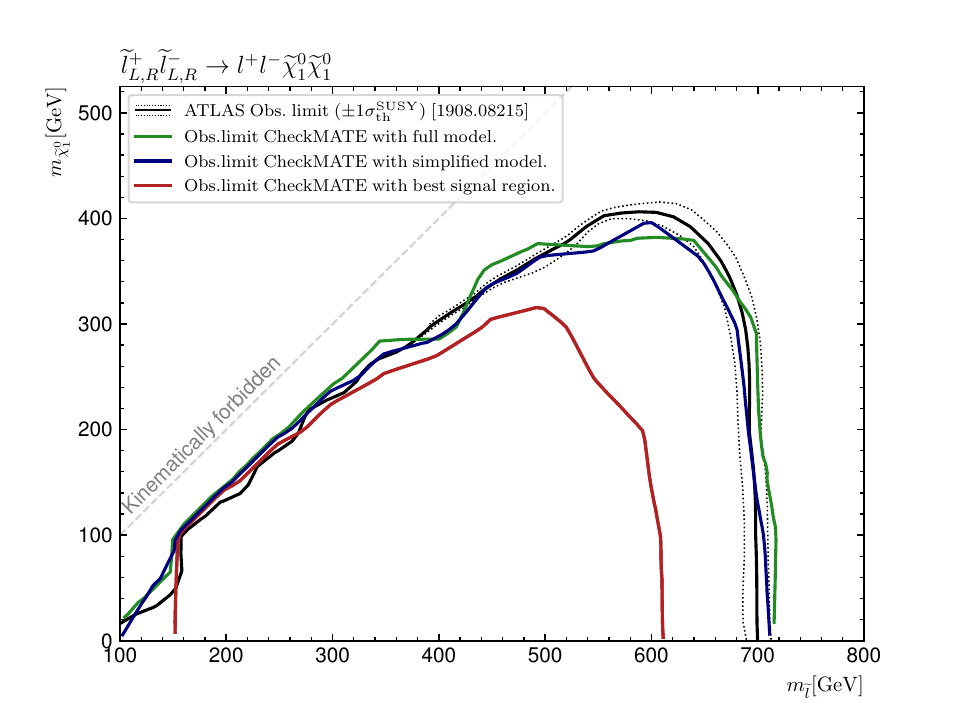}
\includegraphics[width=0.5\textwidth]{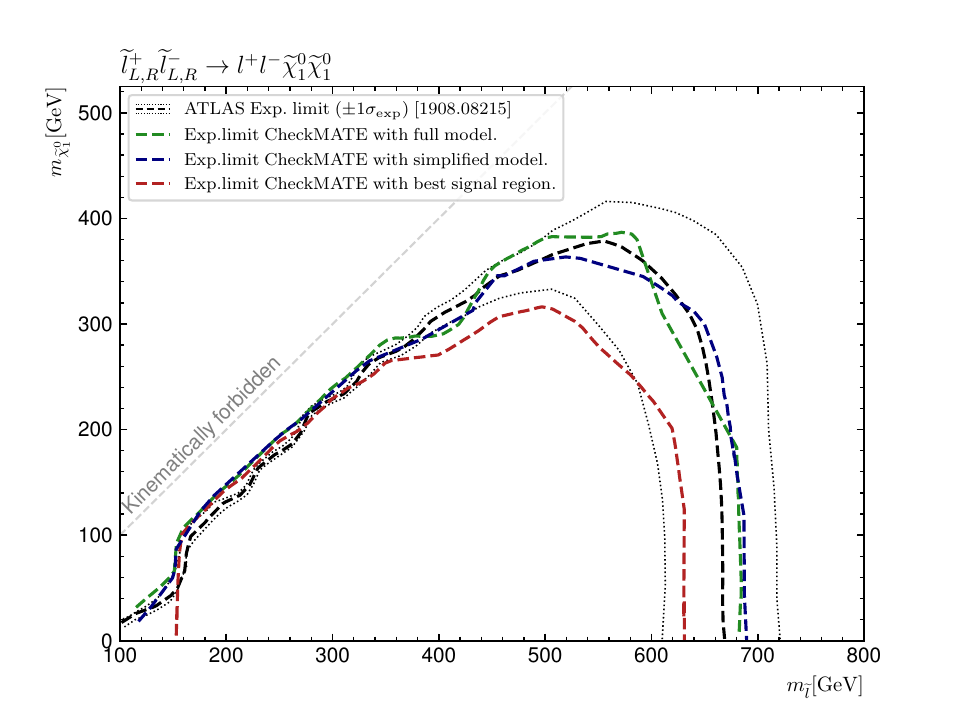}
\caption{Validation plots for the search \texttt{atlas\_1908\_08215} (SUSY-2018-33), slepton pair production. Left panels: observed limits; right panels: expected limits.\label{fig:190808215}}
\end{figure}

\subsection{\texttt{atlas\_1911\_06660} (SUSY-2018-04)}
This is a search \cite{ATLAS:2019gti} for the production of staus in final states with two hadronic $\tau$-leptons and missing transverse momentum. Two orthogonal signal regions can be combined in a fit for which the full likelihood model is provided. The implementation includes one of the multijet control regions (CR-A), for which a significant contribution is expected, up to 30\%, from signal events.

The validation plots in Fig.~\ref{fig:191106660} show a similar sensitivity of the best signal region, as well as full and simplified likelihoods. In each case, good agreement with the ATLAS result is observed. The observed exclusion limits slightly exceed the ATLAS limit, whereas for the expected limits, the best agreement is obtained for the full-likelihood approach, and comfortably within the experimental $1\sigma$ uncertainty. A comparison of observed 95\% CL upper limits on cross section for several representative parameter points is shown in Tab.~\ref{tab:191106660_ul}. In the bottom plot of Fig.~\ref{fig:191106660} we additionally compare the expected limits obtained with and without the inclusion of control regions. Note that for this search, the signal contribution to the control regions is not negligible. Without control regions, the search over-constrains the parameter space, although the discrepancy remains within the 1-$\sigma$ expected uncertainty.

\begin{figure}
\begin{center}
    
\includegraphics[width=0.49\textwidth]{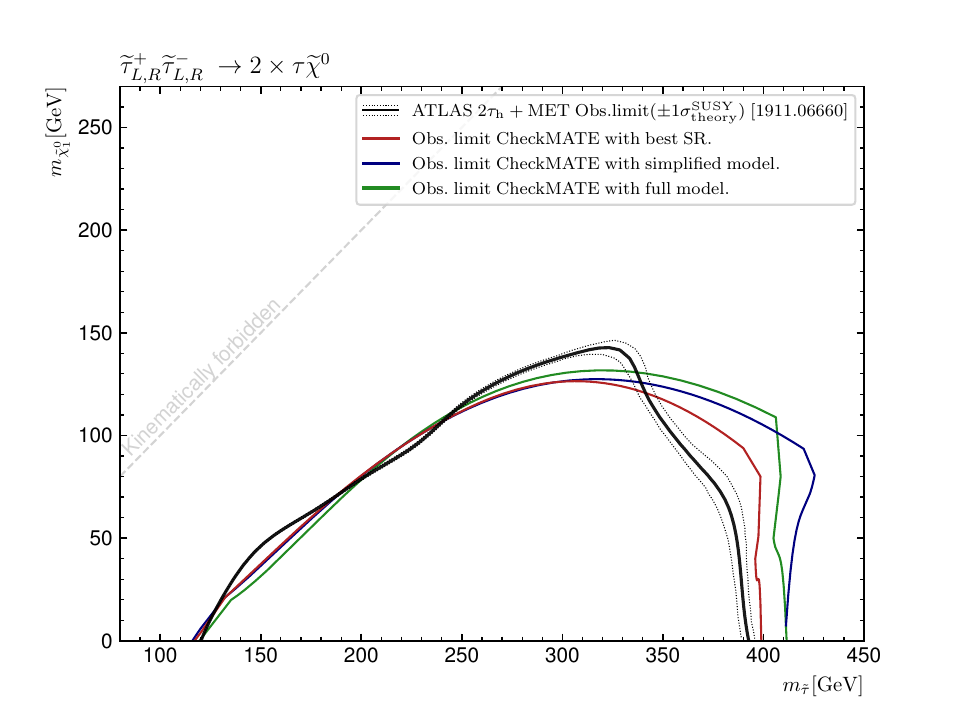}
\includegraphics[width=0.49\textwidth]{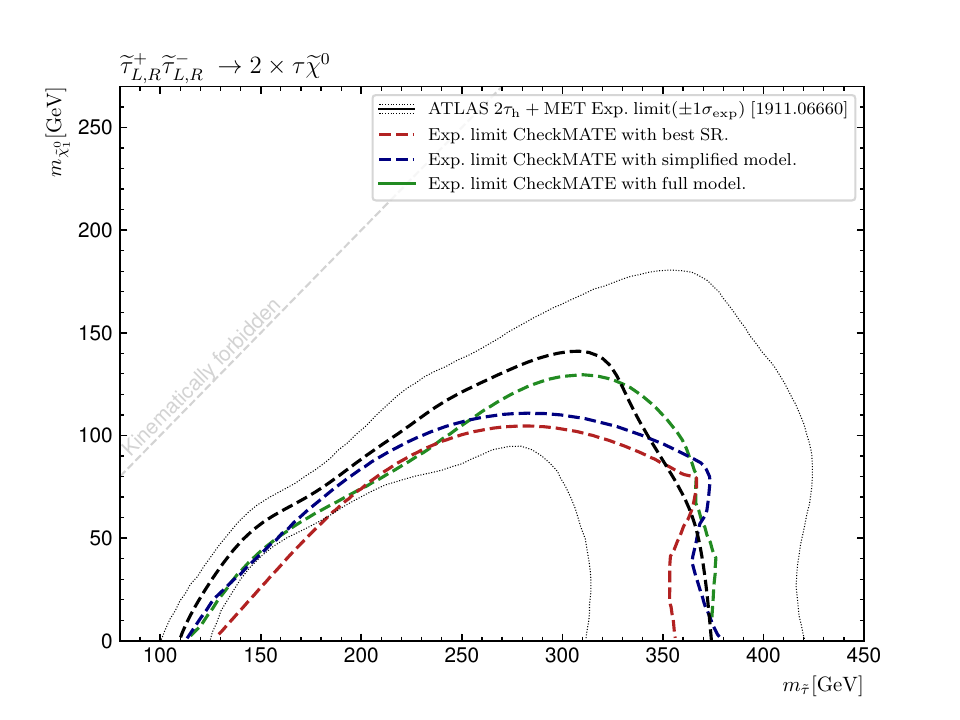}\\
\includegraphics[width=0.5\textwidth]{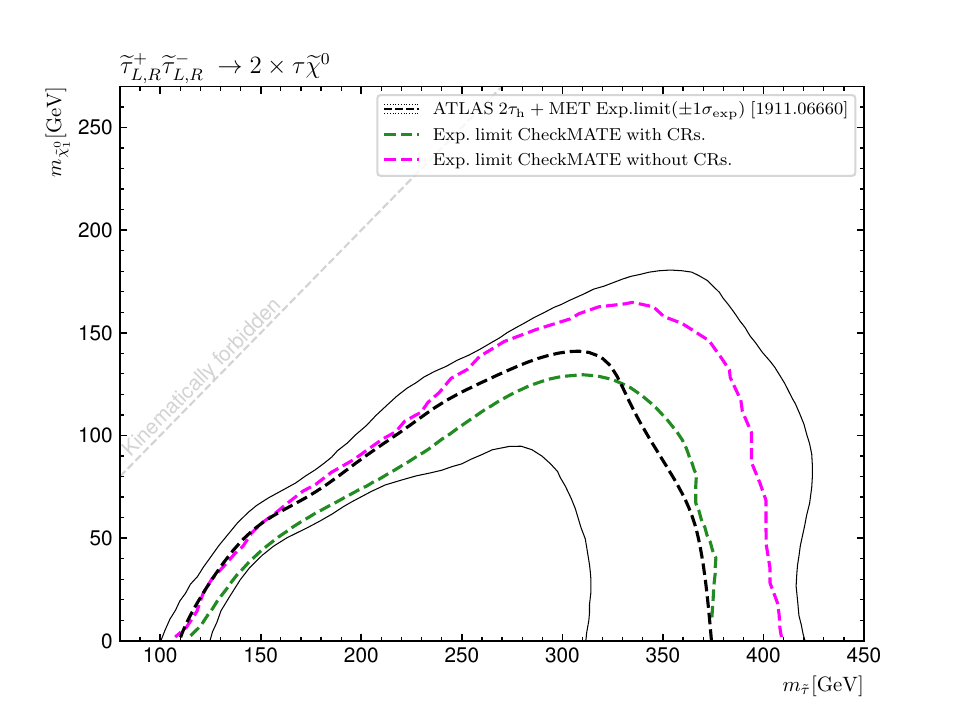}
\end{center}
\caption{Validation plots for the search \texttt{atlas\_1911\_06660} (SUSY-2018-04). Left panels: observed limits; right panels: expected limits. In the bottom row: a comparison between the observed limits derived with and without control regions.\label{fig:191106660}}
\end{figure}

 \begin{table}
  \begin{center}
   \begin{tabular}{l r r r }
\toprule
$(m_{\tilde{\tau}_1}, m_{\tilde{\chi}_1^0})$ [GeV] & full & simple  & ATLAS  \\ \midrule
$ (320,120)$ & $3.7$ & $4.0$ & $4.6$   \\
$ (320,160)$ & $5.2$ & $5.8$ & $5.0$   \\
$ (400,80)$ & $1.9$ & $1.6$ & $2.0$  \\
$ (440,80)$ & $1.5$ & $1.4$ & $1.7$  \\ \bottomrule
\end{tabular}
\caption{Comparison of observed 95\% CL upper limits on cross section in fb for stau pair production in the search \texttt{atlas\_1911\_06660} (SUSY-2018-04); full and simple columns correspond to the limits obtained with \CM{} using the full and simplified likelihood, respectively.\label{tab:191106660_ul}}
  \end{center}
 \end{table}

\subsection{\texttt{atlas\_1911\_12606} (SUSY-2018-16)}
This is a search \cite{ATLAS:2019lng} for the production of electroweakinos and sleptons in scenarios with compressed mass spectra. The final states contain two low $p_T$ leptons (opposite sign and same or different flavor). Sensitivity of the search relies on additional initial state radiation (ISR) jets which give transverse boost to the final-state particles and add missing transverse momentum. There are two multibin signal regions implemented in \CM{}:\footnote{ vector boson fusion (VBF) SRs are currently not available.} \texttt{SR-EWK} targeting electroweakinos production and divided into 44 bins according to the lepton pair invariant mass, \met, and lepton flavor; \texttt{SR-S} targets the production of sleptons and is divided into 32 bins. The full likelihood model is provided. 

In Figure~\ref{fig:191112606}, we show validation plots for slepton search signal regions. The electron and muon channels are shown separately. Very good agreement between ATLAS and \CM{} is obtained for the shape-fit analysis. Additionally, by comparing red exclusion curves obtained using the best-SR-method, we find a clear benefit of using multibin signal regions. Since there is a negligible difference between expected and observed results (both for ATLAS and \CM), we only show the observed limits.

In Figure~\ref{fig:191112606_wino}, we show validation plots for the production of wino-like chargino-neutralino pairs (EW search region). The searches for different parities of LSP and NLSP are shown separately: upper row $m(\tilde{\chi}_{2}^{0}) \times m(\tilde{\chi}_{1}^{0})>0$ (same parity) and $m(\tilde{\chi}_{2}^{0}) \times m(\tilde{\chi}_{1}^{0})<0$ (opposite parity) in the lower row; see, e.g.\ Ref.~\cite{Choi:2005gt} for a discussion of CP properties of the neutralino sector and consequences for the decay kinematics. Clearly, the simplified likelihood has a problem with reproducing the complicated shape of the observed limit due to significant excesses and deficits in several signal regions. Note that a similar conclusion was obtained by ATLAS in Ref.~\cite{ATL-PHYS-PUB-2021-038}. Therefore, for this search, we advise against using the simplified likelihood in routine runs. The agreement between the full likelihood and the ATLAS observed limit for the case of same parity neutralinos (top row) is not ideal as well, even though the expected limits comparison is rather encouraging and within 1-$\sigma$ uncertainty, see the bottom plot in Fig.~\ref{fig:191112606_wino}. The search is also included in \textsc{HackAnalysis}, cf.\ Ref.~\cite{Agin:2024yfs}, \textsc{SModelS}~\cite{Altakach:2024jwk}, and \textsc{MadAnalysis}~\cite{Araz:2025bww}. Additional validation material for \CM, including cut-flows, can be found in~\cite{191112606_validation}.

\begin{figure}
\includegraphics[width=0.5\textwidth]{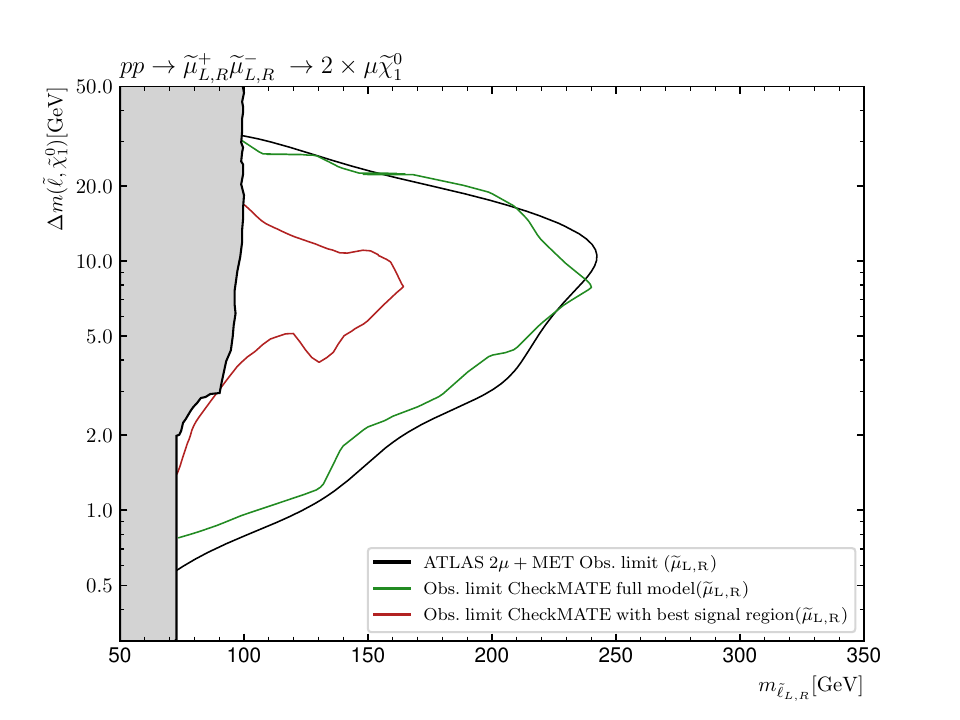}
\includegraphics[width=0.5\textwidth]{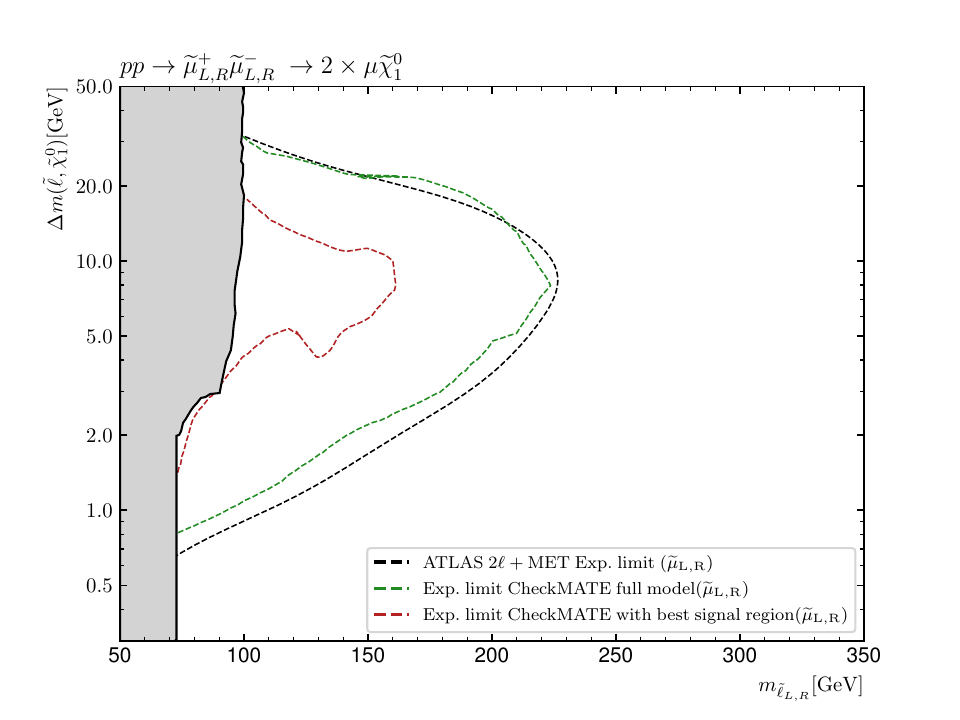}\\
\includegraphics[width=0.5\textwidth]{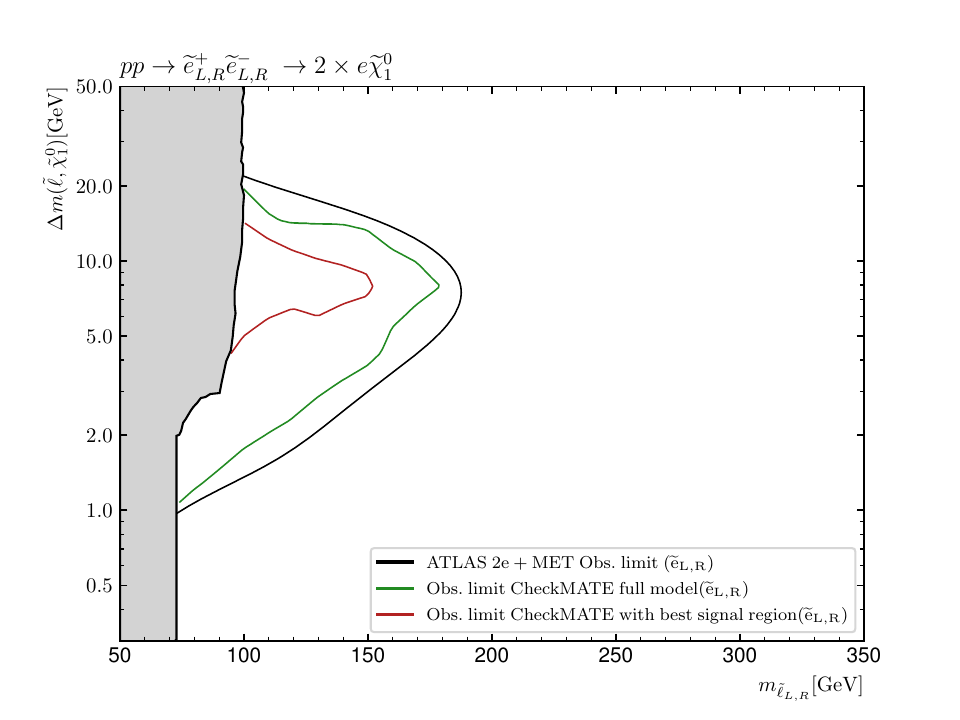}
\includegraphics[width=0.5\textwidth]{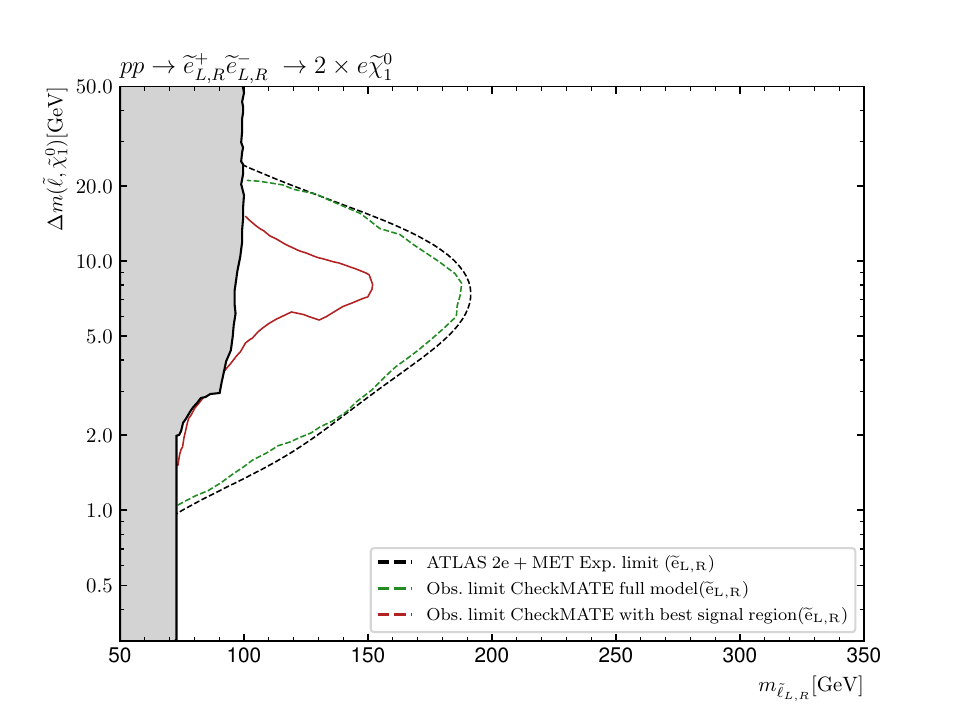}
\caption{Validation plots (left: observed limits; right: expected limits) for the search \texttt{atlas\_1911\_12606} (SUSY-2018-16), slepton pair production. Top row: smuon search; bottom row: selectron search. Left panel: observed limits; right panel:  expected limits.
\label{fig:191112606}}
\end{figure}

\begin{figure}
\begin{center}
\includegraphics[width=0.49\textwidth]{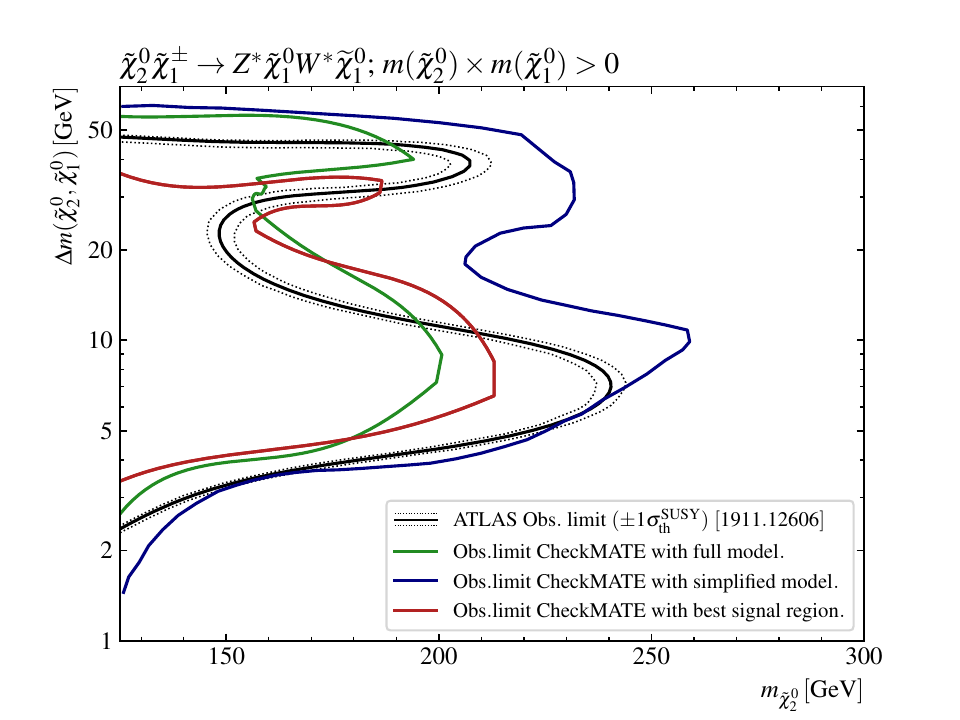}
\includegraphics[width=0.49\textwidth]{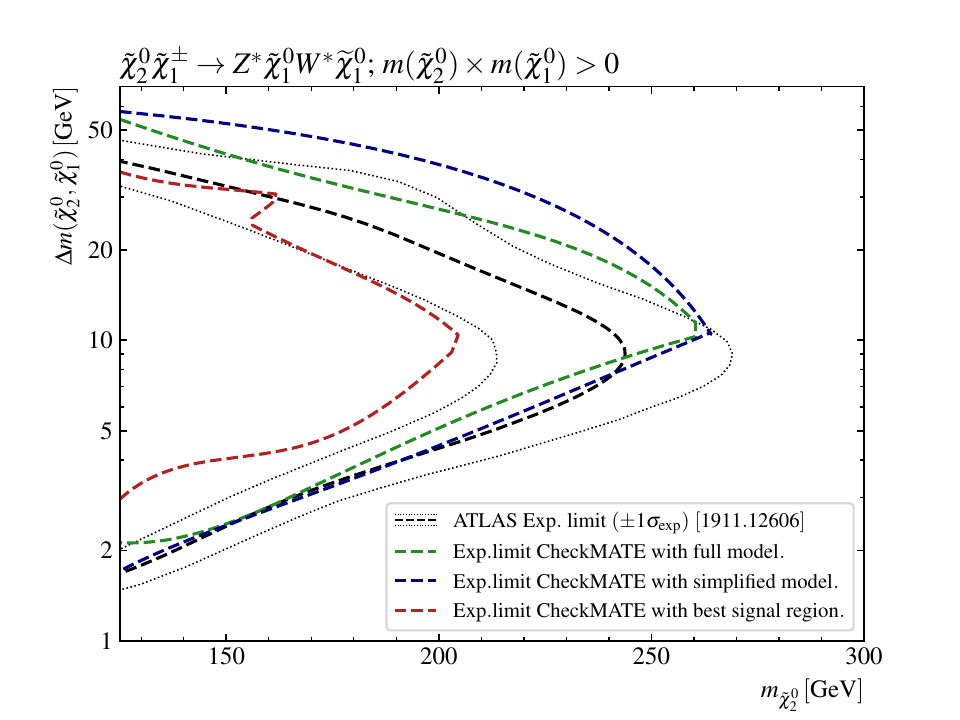}\\
\includegraphics[width=0.49\textwidth]{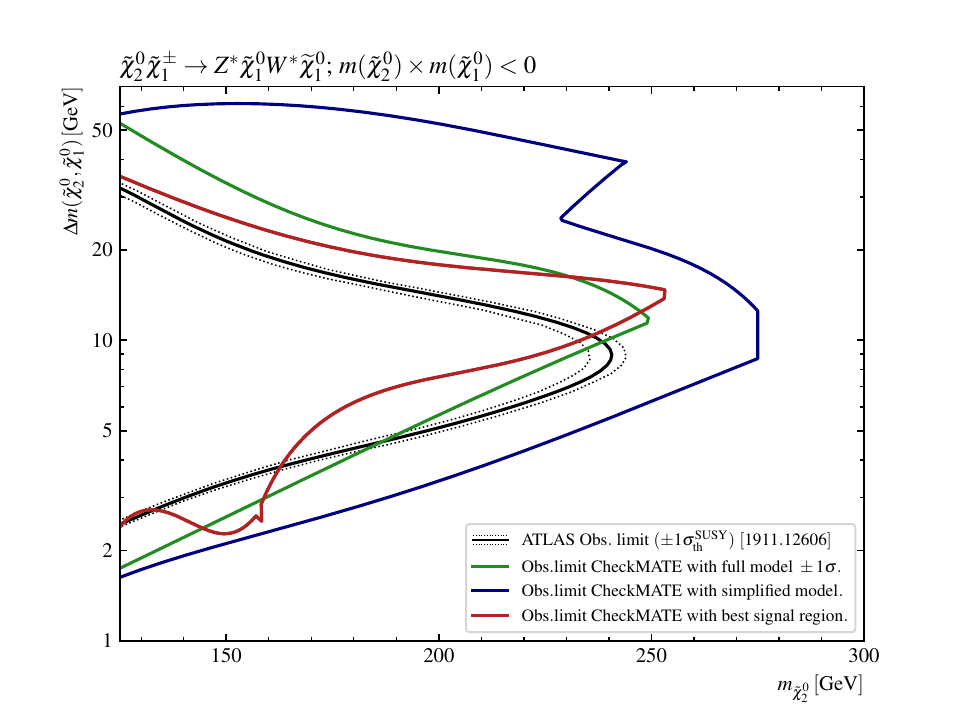}
\includegraphics[width=0.49\textwidth]{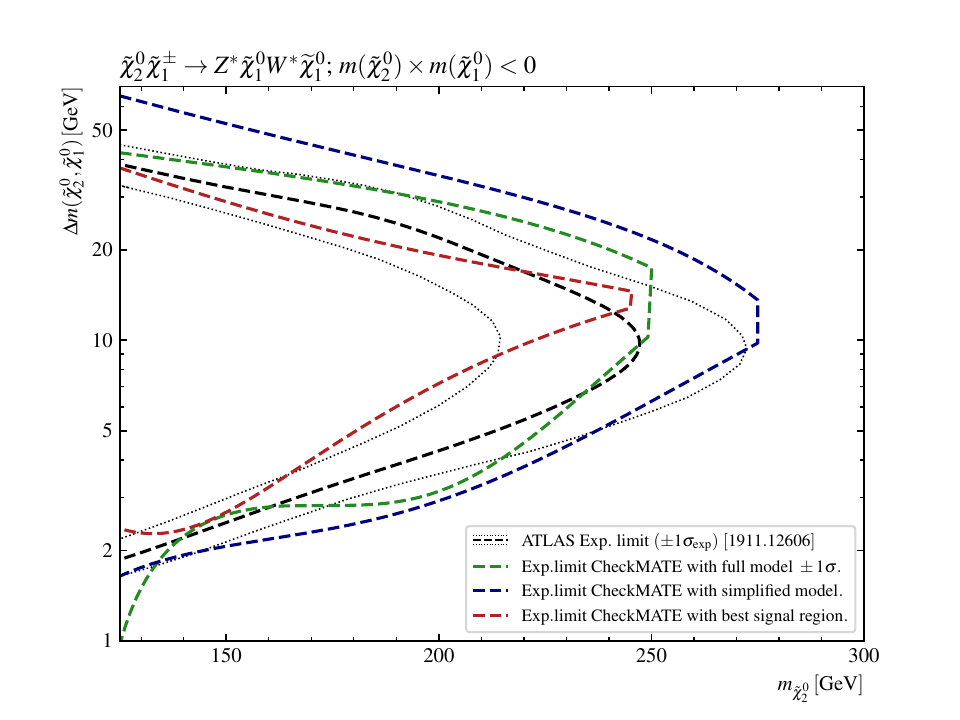}\\
\includegraphics[width=0.49\textwidth]{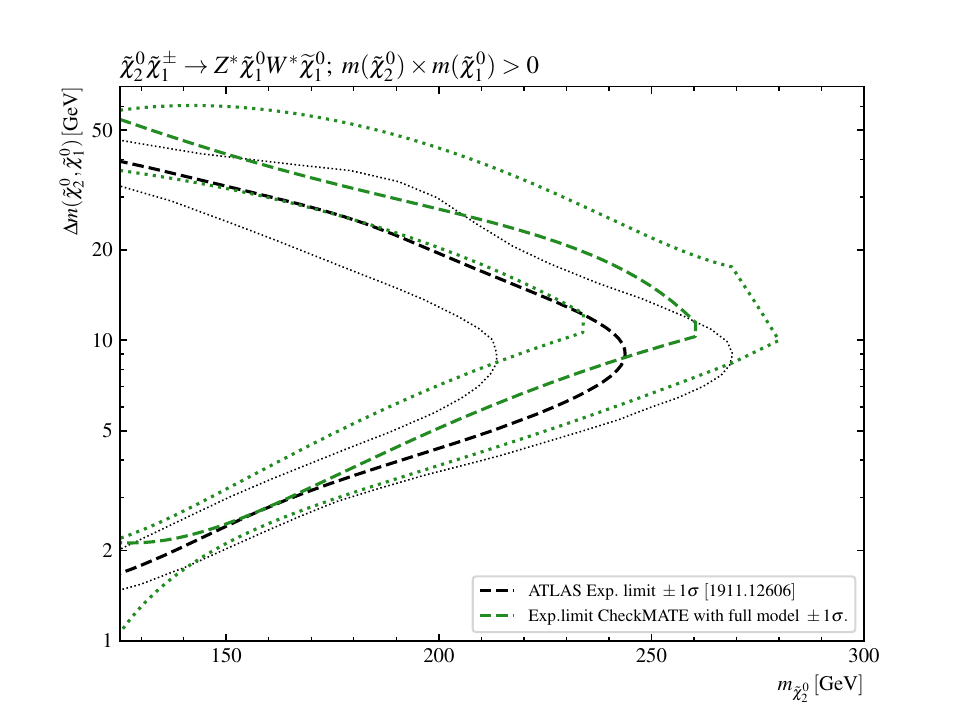}
\end{center}
\caption{Validation plots for the search \texttt{atlas\_1911\_12606} (SUSY-2018-16). Top row: $m(\tilde{\chi}_{2}^{0}) \times m(\tilde{\chi}_{1}^{0})>0$; middle row: $m(\tilde{\chi}_{2}^{0}) \times m(\tilde{\chi}_{1}^{0})<0$. Left panels: observed limits; right panels: expected limits. In the bottom row comparison of expected 1-$\sigma$ bands of ATLAS and \CM{} full likelihood model for the same-parity case. \label{fig:191112606_wino}}
\end{figure}

\subsection{\texttt{atlas\_2004\_14060} (SUSY-2018-12)}
This is a search~\cite{ATLAS:2020dsf} for hadronically decaying supersymmetric partners of top quark, and up-type, 3rd generation scalar leptoquark. The final states consist of several jets, 0 leptons, and large \met. The requirements for $b$ jets vary between signal regions. There are 3 multibin signal regions: \texttt{SRA-B} that targets scenarios with highly boosted top quarks in the final state and is divided into 6 bins according to an invariant mass of a large-$R$ jet; \texttt{SRC}, for scenarios with 3-body decays of stops, that is divided into 5 bins according to a recursive jigsaw reconstruction technique variable $R_\textrm{ISR}$~\cite{Jackson:2016mfb}; \texttt{SRD} for scenarios with compressed spectra and 4-body decays of stops which is divided into 3 bins according to the number of identified $b$-jets. No likelihood model is provided, and only simplified likelihood fitting is available. The combinations are performed separately for each signal category (corresponding to three likelihood models), A-B, C, and D, as described in the ATLAS paper. This avoids overlaps and accidental correlations. 

The validation plots for this search are shown in Figure~\ref{fig:200414060}. We include the full range of stop masses and decay modes. Good agreement across the parameter plane is observed. For the high-stop-mass region the \CM{} result is slightly weaker than ATLAS, clearly extending the exclusion reach compared to single SR exclusion. In more compressed scenarios, the simplified fit gives a slightly too strong exclusion in some regions of the parameter space; however, this is below 30\% difference in the exclusion strength.  Additional validation material can be found in~\cite{200414060_validation}.  

\begin{figure}
       \includegraphics[width=0.5\textwidth]{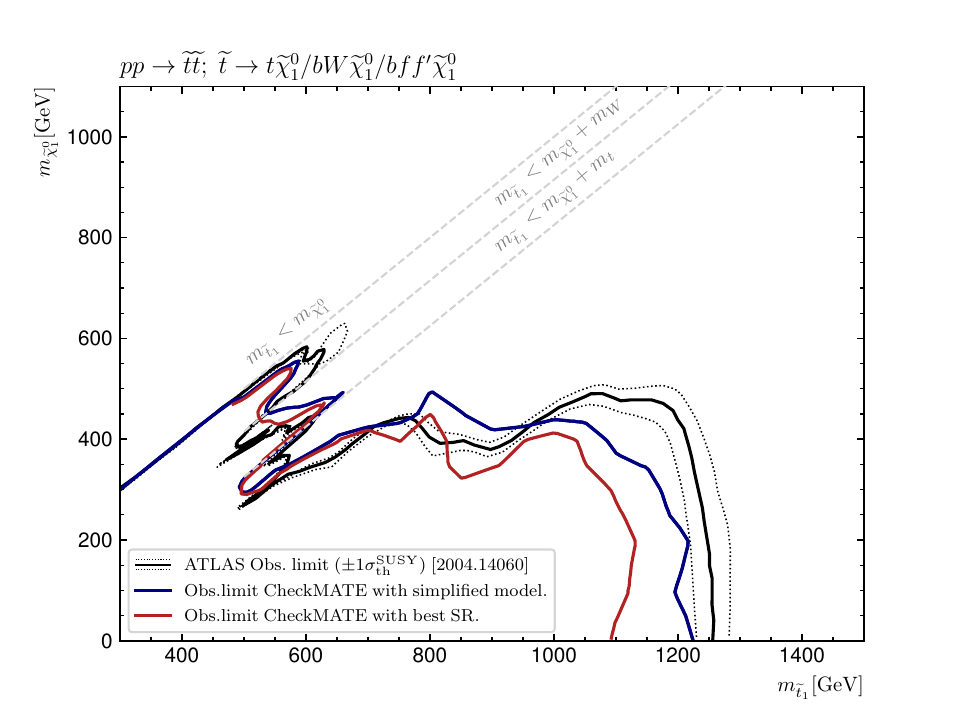}
 \includegraphics[width=0.5\textwidth]{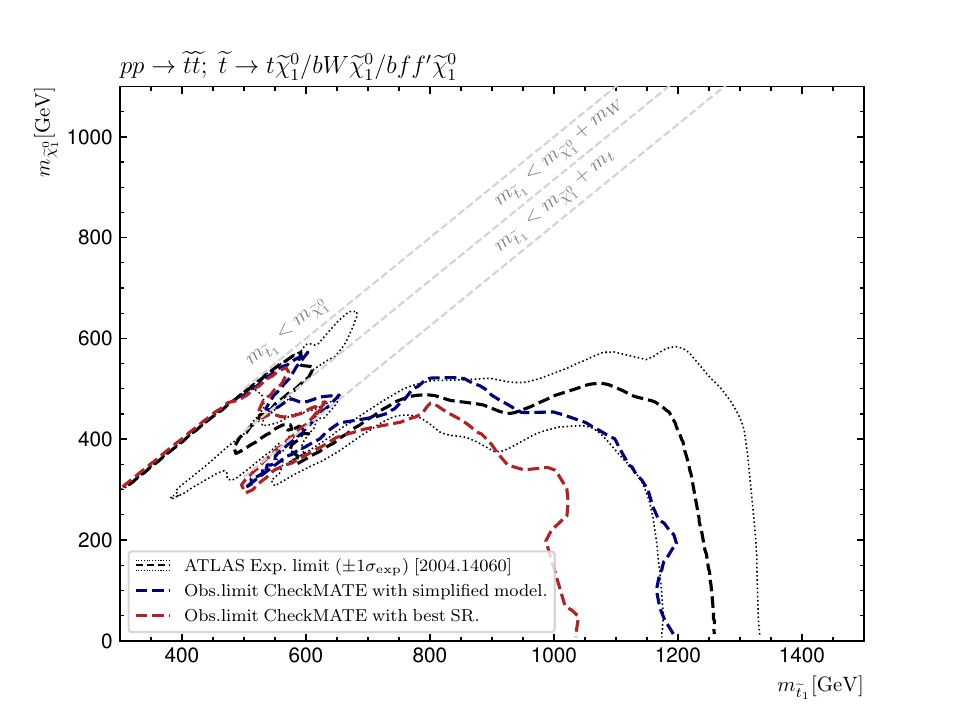}
    \caption{Validation plots for the search \texttt{atlas\_2004\_14060} (SUSY-2018-12). Left panels: observed limits; right panels: expected limits.
    \label{fig:200414060}}
\end{figure}

\subsection{\texttt{atlas\_2006\_05880} (SUSY-2018-21)}
This is a search~\cite{ATLAS:2020aci} for supersymmetric partners of top quark decaying to a Higgs or $Z$ boson. The final state Higgs boson is reconstructed from a pair of $b$-jets, whereas the $Z$ boson is reconstructed from a same-flavor opposite-sign dilepton pair. There are 3 multibin signal regions that are shape-fits in \met, \met-significance, and $p_\T$ of the $Z$ candidate. No likelihood model is provided, and only a simplified likelihood model is available. The combinations are performed separately for three signal categories: SR$^{hZ}_{\mathrm{1A}}$ (combining signal regions SR{$_{\mathrm{1A}}^Z$}, SR{$_{\mathrm{1A}}^h$}, SR{$_{\mathrm{1AB}}^h$}), SR$^{hZ}_{\mathrm{1B}}$ (combining signal regions SR{$_{\mathrm{1B}}^Z$}, SR{$_{\mathrm{1B}}^h$}, SR{$_{\mathrm{1AB}}^h$}), SR$^Z_2$ (combining signal regions SR{$_{\mathrm{2A}}^Z$} and SR{$_{\mathrm{2B}}^Z$}).   This ensures statistical independence of signal regions in each combination and follows ATLAS definitions of likelihood models.

The validation plots in Figure~\ref{fig:200605880} again show a good agreement, with \CM{} somewhat weaker than ATLAS in the high stop-mass region. The observed limits clearly extend the reach compared to the best SR, increasing sensitivity by up to factor 2.

\begin{figure}
    \includegraphics[width=0.5\textwidth]{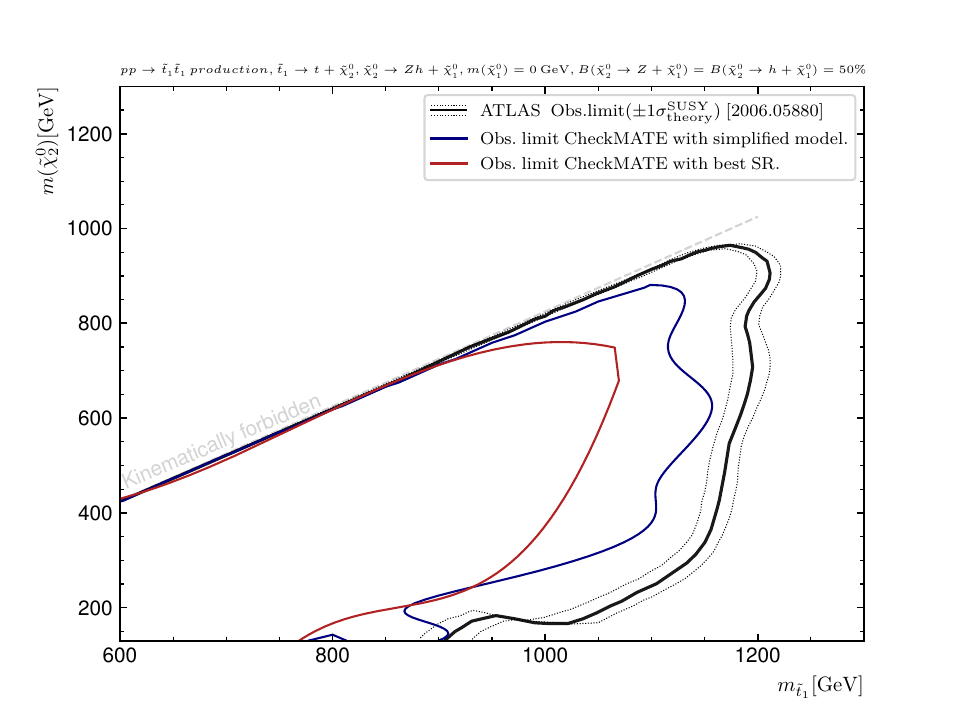}
 \includegraphics[width=0.5\textwidth]{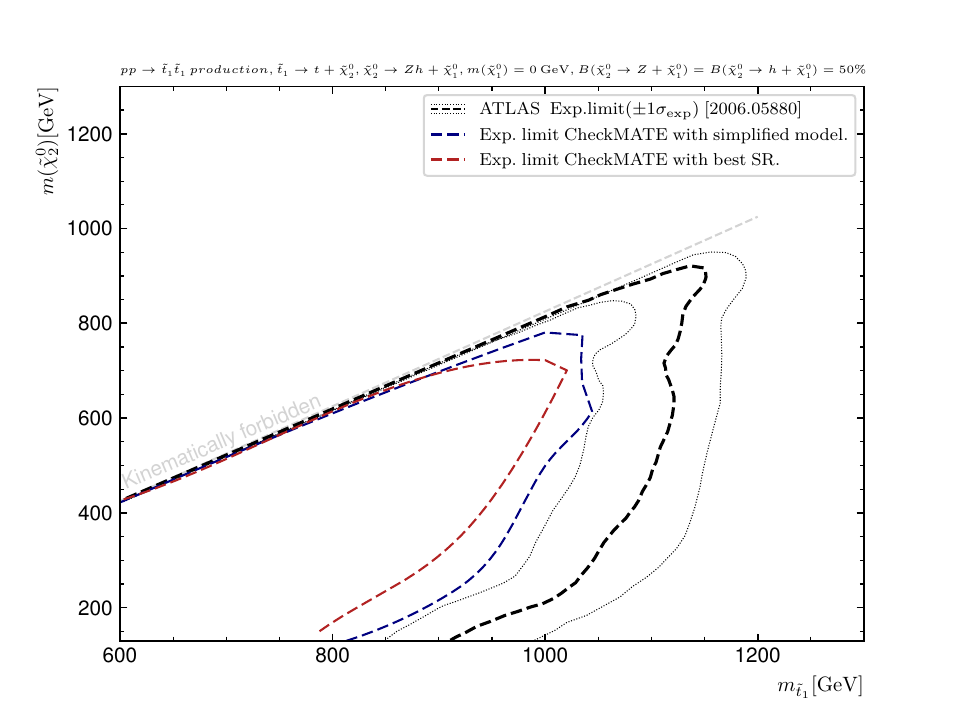}
 \caption{Validation plots for the search \texttt{atlas\_2006\_05880} (SUSY-2018-21). Left panels: observed limits; right panels: expected limits.
    \label{fig:200605880}}
\end{figure}

\subsection{\texttt{atlas\_2010\_14293} (SUSY-2018-22)}
This is a search \cite{ATLAS:2020syg} for squarks (1st and 2nd generation) and gluinos in final states with 2-6 jets, 0 leptons, and missing transverse momentum. There are three multibin signal regions in this search: \texttt{MB-SSd} which targets squark pair production and is divided into 24 bins according to $m_\textrm{eff}$, $\met/\sqrt{H_\T}$ and the number of jets; \texttt{MB-GGd} which targets gluino production and is divided into 18 bins according to $m_\textrm{eff}$, $\met/\sqrt{H_\T}$; \texttt{MB-C} which targets compressed spectra and is divided into 18 bins according to $m_\textrm{eff}$, $\met/\sqrt{H_\T}$ and the number of jets. The control regions are also implemented, which gives an opportunity to fully exploit the full likelihood model provided by the ATLAS Collaboration.

Figure~\ref{fig:201014293} shows the validation plots for the squark pair-production process. The simplified shape-fit has excellent agreement with the official result, while the full likelihood model gives a somewhat weaker exclusion. In either case, there is a clearly visible advantage over the best-SR method. In Appendix we provide detailed signal yields for the multibin SRs of this search.  Additional validation material (cut flows) can be found in~\cite{201014293_validation}.

\begin{figure}
    \includegraphics[width=0.5\textwidth]{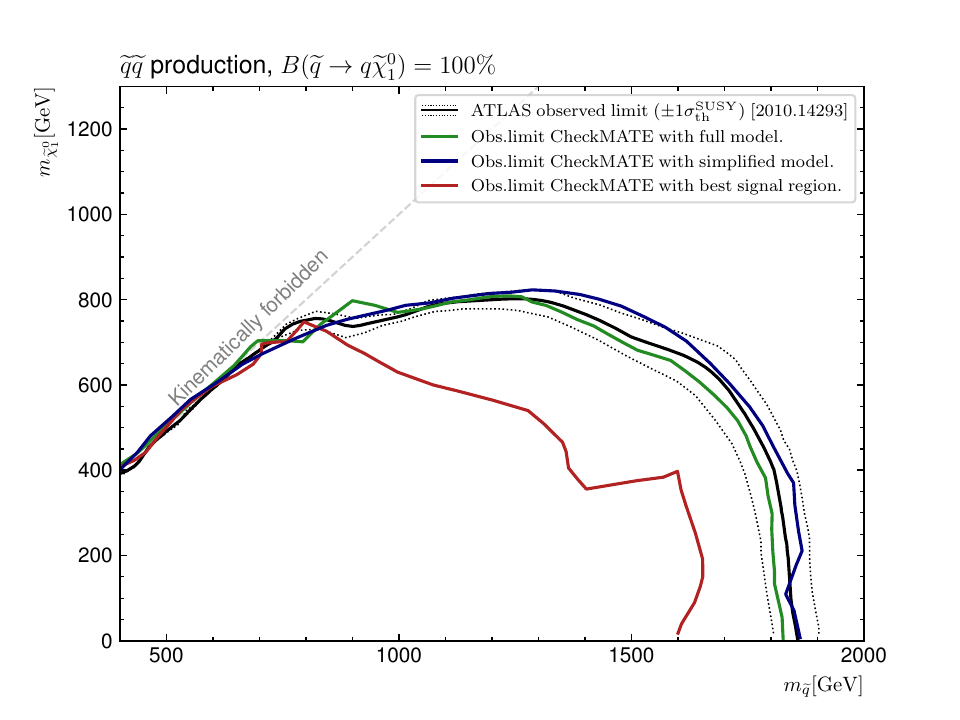}
 \includegraphics[width=0.5\textwidth]{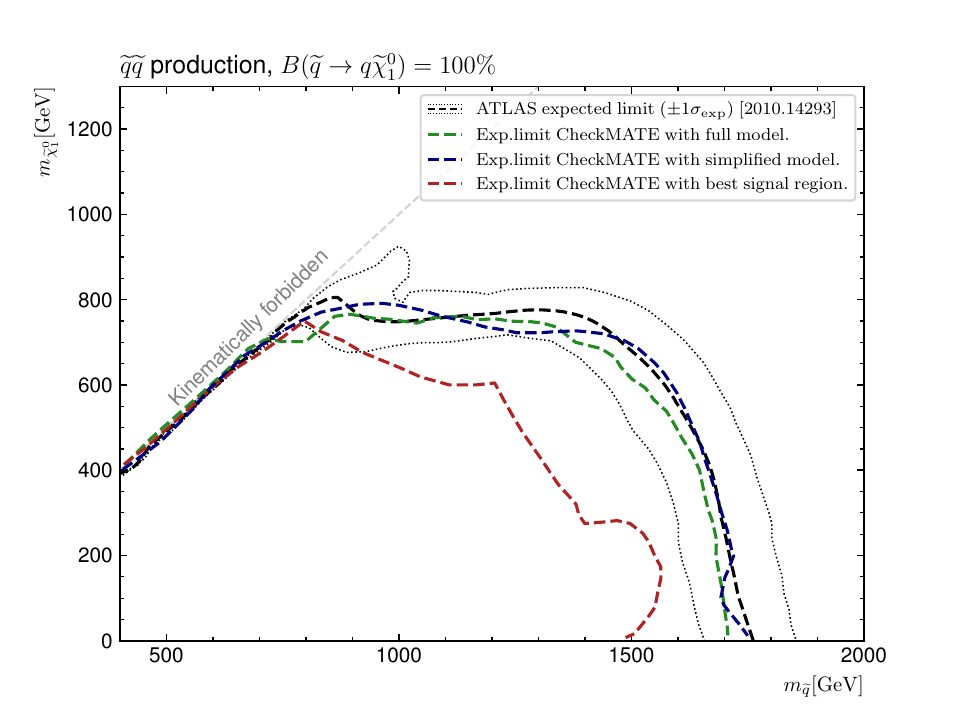}
    \caption{Validation plots for the search \texttt{atlas\_2010\_14293} (SUSY-2018-22), squark pair production - 8-fold degeneracy. Left panels: observed limits; right panels: expected limits.
    \label{fig:201014293}}
\end{figure}

\subsection{\texttt{atlas\_2101\_01629} (SUSY-2018-10)}
This is a search \cite{ATLAS:2021twp} for squarks and gluinos in final states with an isolated lepton, jets, and missing transverse momentum. Benchmark models assume long decay chains for squarks and gluinos with charginos, neutralinos and gauge bosons in intermediate states that give rise to the final state lepton. There is a multibin signal region that combines 26 bins defined according to the number of jets, the number of identified $b$-jets, and $m_\textrm{eff}$. The full likelihood model is provided. 

The validation plots for gluino pair production, followed by the decay to intermediate mass chargino, are shown in Fig.~\ref{fig:210101629}. The shape of the ATLAS exclusion line is well reproduced for the likelihood models, whereas the BSR exclusion is significantly weaker. 

\begin{figure}
      \includegraphics[width=0.5\textwidth]{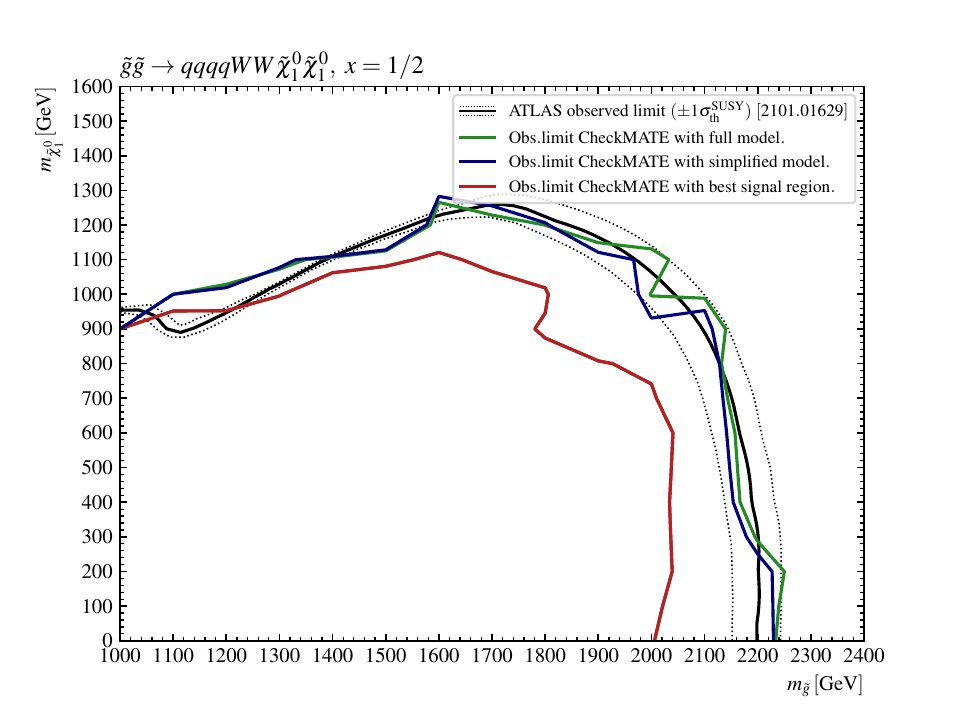}
 \includegraphics[width=0.5\textwidth]{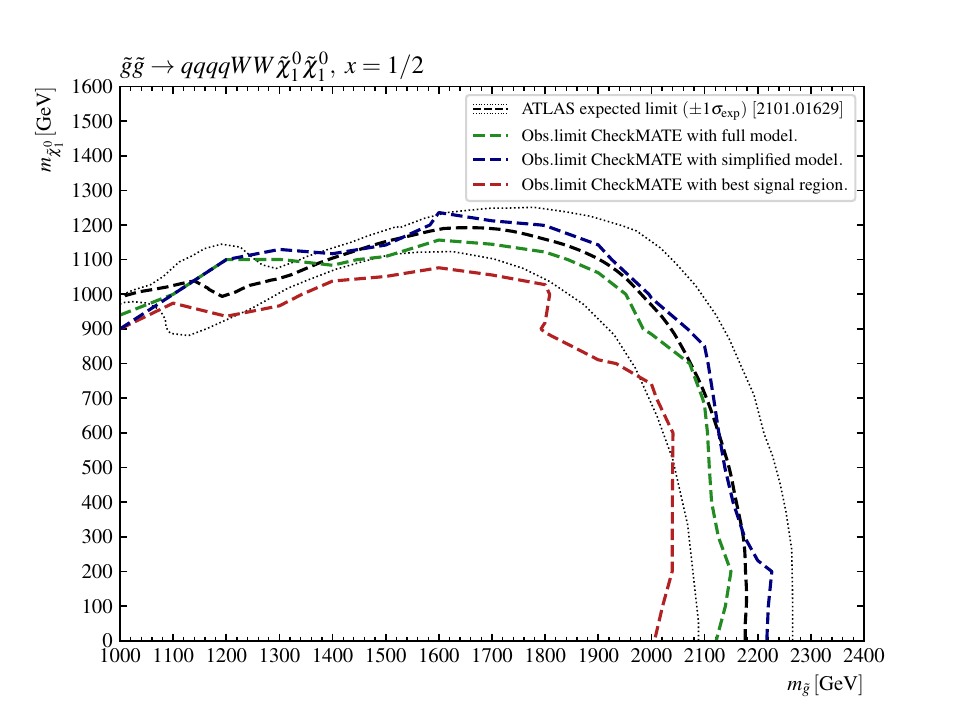}
 \caption{Validation plots for the search \texttt{atlas\_2101\_01629}, gluino pair production and the decay chain: $\tilde{g}\to q\bar{q}' \tilde{\chi}_1^\pm\to q\bar{q}'W \tilde{\chi}_1^0 $ with $x = (m(\tilde{\chi}_1^\pm) - m(\tilde{\chi}_1^0))/(m(\tilde{g}) - m(\tilde{\chi}_1^0)) = 1/2$.\label{fig:210101629} }
   \end{figure}

\subsection{\texttt{atlas\_2111\_08372} (HIGG-2018-26)}
This is a search \cite{ATLAS:2021gcn} for invisible decays of a Higgs boson or dark matter (DM) particles, $\chi$, produced in association with a $Z$ boson. The final state $Z$ boson is reconstructed from a same-flavor opposite-sign dilepton pair. A multibin signal region, optimized for the 2HDM+$a$ model~\cite{Bauer:2017ota}, is implemented in \CM{}. It is divided into 22 bins according to the transverse mass, $m_\T$. The implementation is based on the simplified likelihood model.  

As can be seen in Figure~\ref{fig:211108372} we find excellent agreement between the simplified shape fit result and the official result. The best-SR exclusion is significantly weaker. It should be noted that in the parts of the plot with large $m_a$ and/or $m_A$ the exclusion is driven by the shape fit in the range of large $m_T$, which increases the sensitivity by factor $\sim2$.  

\begin{figure}
\includegraphics[width=0.5\textwidth]{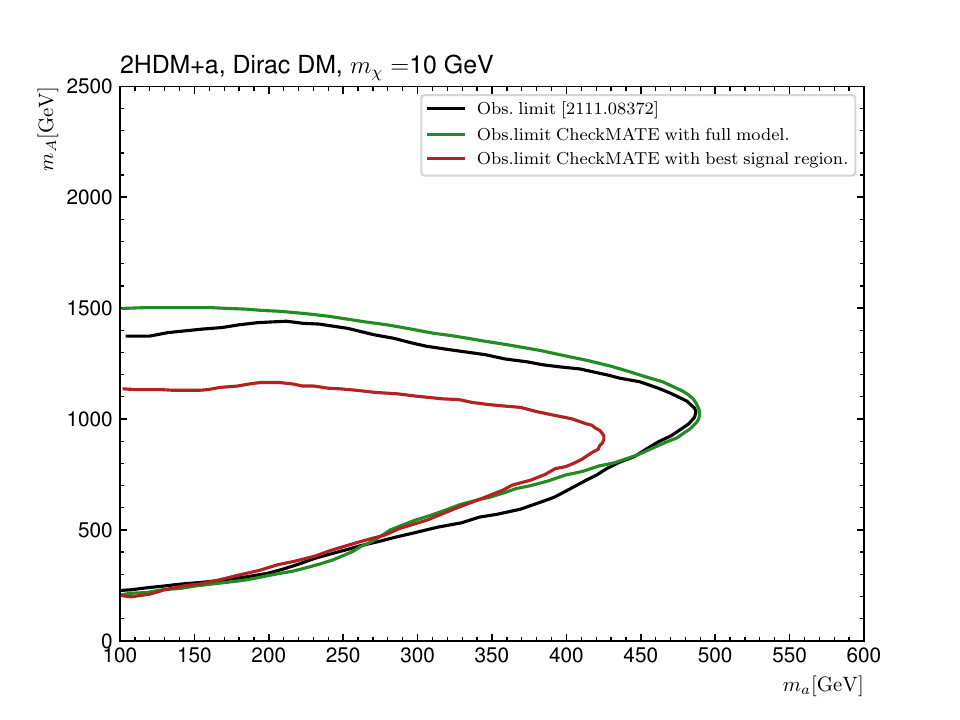}
 \includegraphics[width=0.5\textwidth]{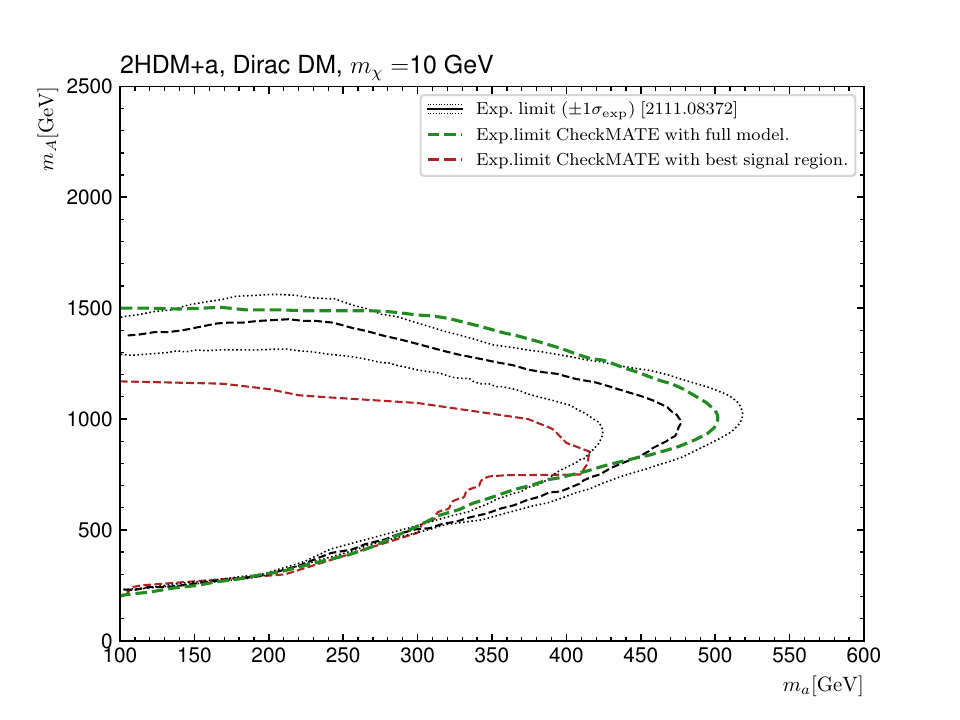}
\caption{Validation plots for the search \texttt{atlas\_2111\_08372} (HIGG-2018-26) in the 2HDM+$a$ model, $pp \to Z(\to \ell^+\ell^-) a(\to \chi\chi)$. Left panel: observed limits; right panel: expected limits.
    \label{fig:211108372}}
\end{figure}

\subsection{\texttt{cms\_1908\_04722} (SUS-19-006)}
This is a search \cite{CMS:2019zmd} for supersymmetric particles in final states with jets ($\geq 2$) and missing transverse momentum. The main discriminating variables are $H_T$ and $H_\T^\mathrm{miss}$, the scalar sum of jet transverse momenta, and the magnitude of the vector sum of jet transverse momenta, respectively. Only the signal regions for prompt production are currently implemented in \CM{}. The search combines 174 bins in the simplified likelihood framework with the covariance matrix provided by the CMS Collaboration. Individual bins are defined according to the number of jets and $b$-jets, $H_\T$, and $H_\T^\mathrm{miss}$. In addition, there are 12 aggregate signal regions defined.  

In Figure~\ref{fig:190804722} we provide validation plots for two cases: squark pair production (single generation) and gluino pair production (followed by $\tilde{g} \to t \bar{t} \tilde{\chi}_1^0$ decay). In both cases, we observe good agreement, within 1-sigma uncertainty) between the \CM{} likelihood model and the CMS results. In case of gluinos there is a significant improvement with respect to the best-SR result, especially in the large gluino-mass limit. It is worth noting the small production cross section in this region: $\sigma \sim 0.6$~fb at $m_{\tilde{g}} \sim 2.2$~TeV.

\begin{figure}
      \includegraphics[width=0.5\textwidth]{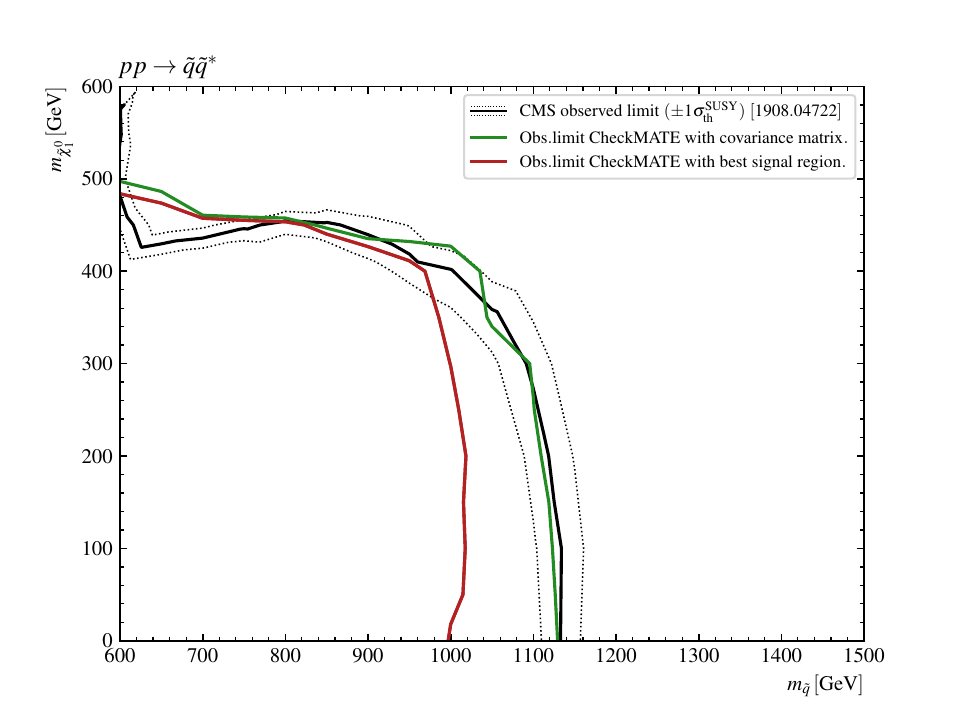}
 \includegraphics[width=0.5\textwidth]{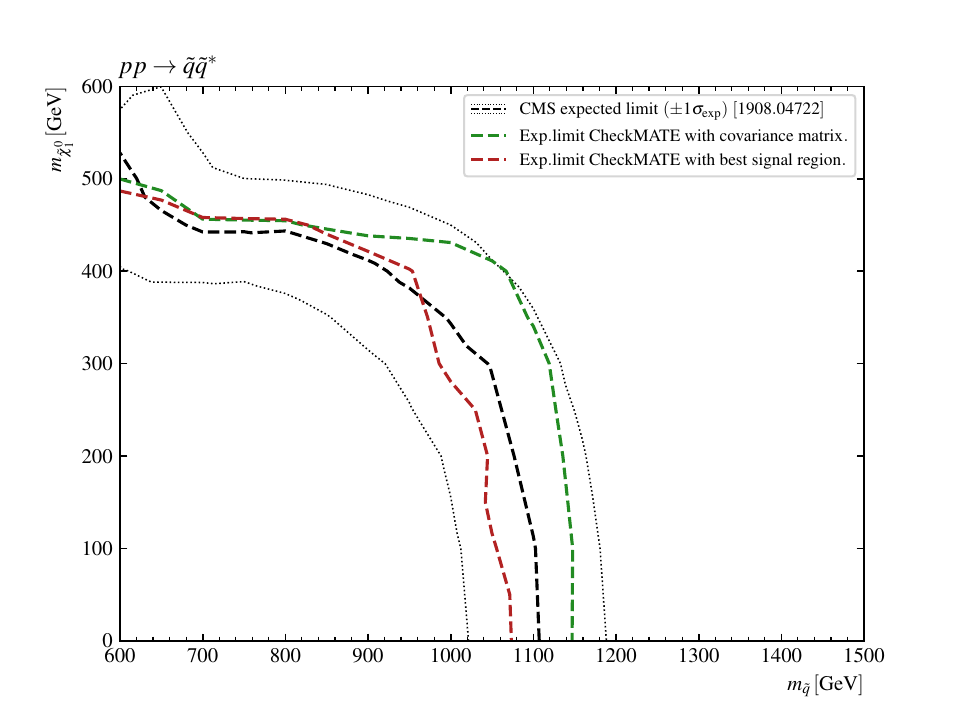}\\
 \includegraphics[width=0.5\textwidth]{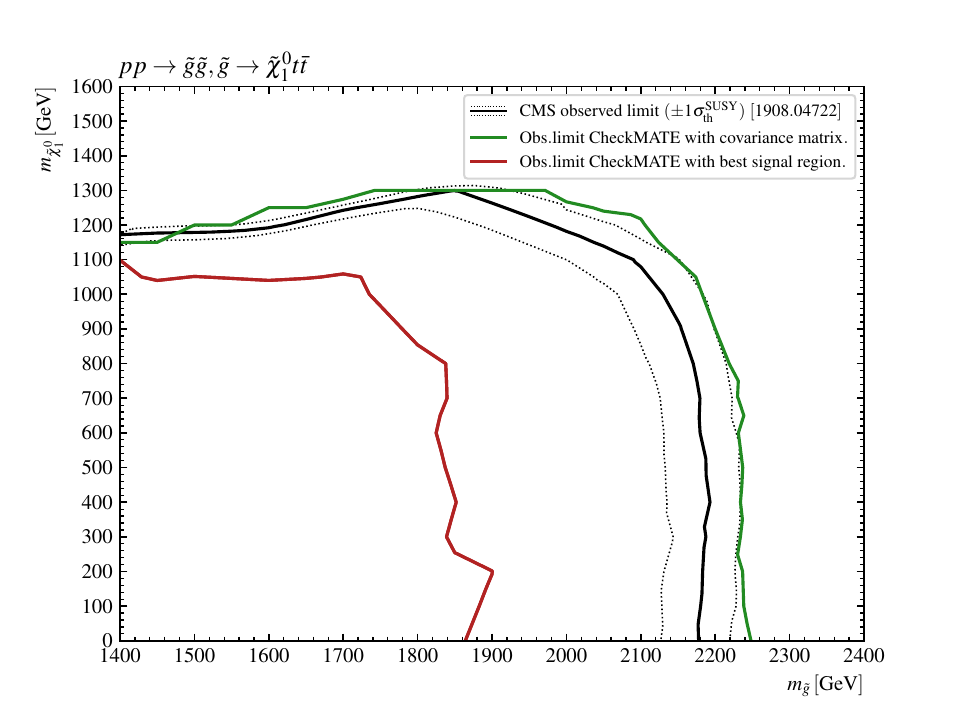}
 \includegraphics[width=0.5\textwidth]{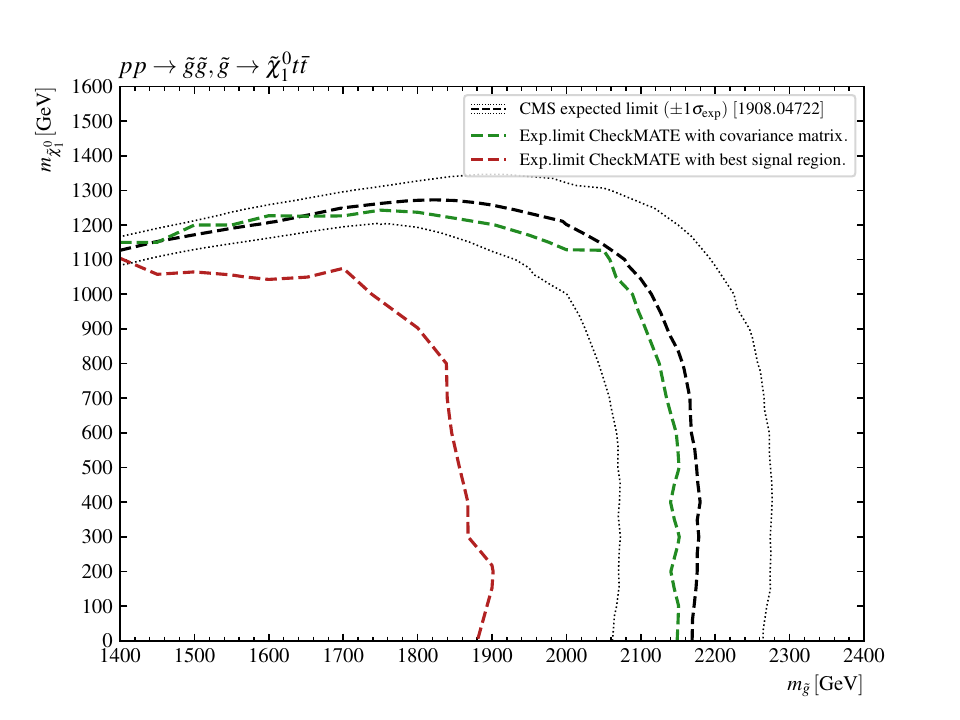}
     \caption{Validation plots for the search \texttt{cms\_1908\_04722}, squark pair production (1-fold degeneracy; upper row), gluino pair production with $\tilde{g}\to t \bar{t} \tilde{\chi}^0_1$ (lower row). The dotted lines around ATLAS limit denote 1-sigma uncertainty: theoretical for the observed limit and experimental for the expected limit. Left panels: observed limits; right panels: expected limits.\label{fig:190804722}} 
   \end{figure}

\subsection{\texttt{cms\_1909\_03460} (SUS-19-005)}
This is a search \cite{CMS:2019zmd} for supersymmetric particles (squarks and gluinos) in final states with jets ($\geq 2$) and a large missing transverse momentum. The events are required to have at least one energetic jet. For the inclusive search, events are required to have at least 2 jets and signal regions are based on the number of jets, $b$-jets, $H_T$, and the $M_{T2}$ variable calculated using two pseudojets. Only the signal regions for prompt production are currently implemented in \CM{}. There are 21 inclusive aggregate signal regions and 282 exclusive bins, which are combined into two multi-bin signal regions using a covariance matrix. Due to numerical problems with the entire matrix $282\times 282$, we split the calculation into two multibin signal regions (corresponding to final states with low and high-$H_T$, respectively) that are marginally correlated~\cite{CMS-com}. However, these two signal regions can be further combined by a user with the \texttt{SRCombination} input parameter, and the \texttt{UnCorrStatisticsCombiner} function of the \spey{} package.

In Figure~\ref{fig:190903460} we show the validation plot for gluino pair production followed by decay $\tilde{g} \to t\bar{t} \tilde{\chi}_1^0$. The result is based on the above-mentioned combination of multibin signal regions. There is good agreement between \CM{} (green lines) and ATLAS (black lines). There is also a clear advantage of the multibin fir over the best-SR (here the aggregate SRs were used).

\begin{figure}
      \includegraphics[width=0.5\textwidth]{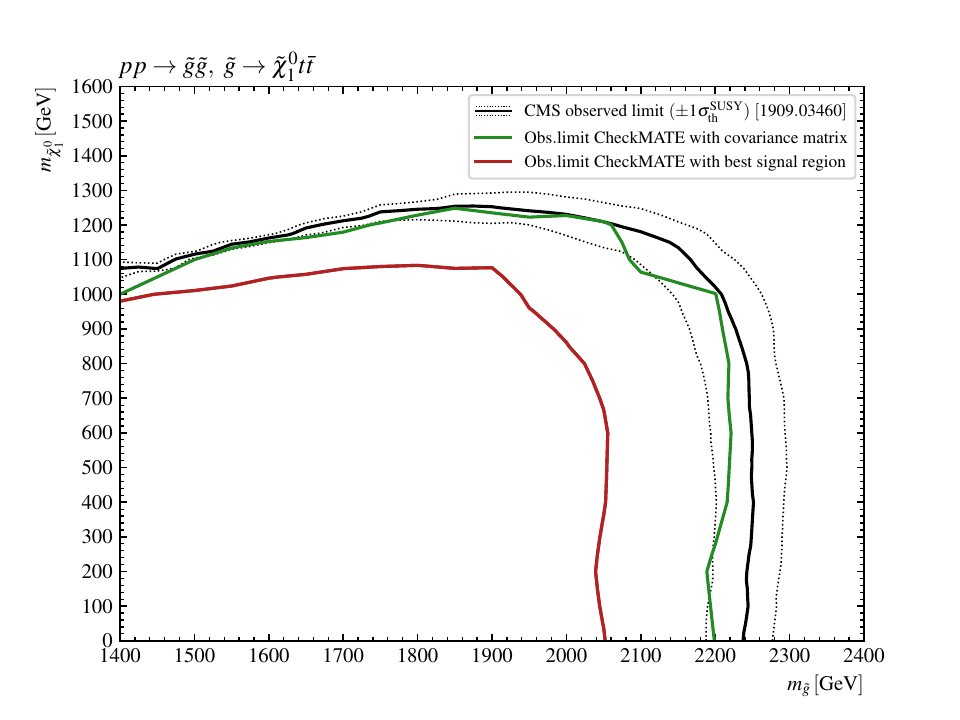}
 \includegraphics[width=0.5\textwidth]{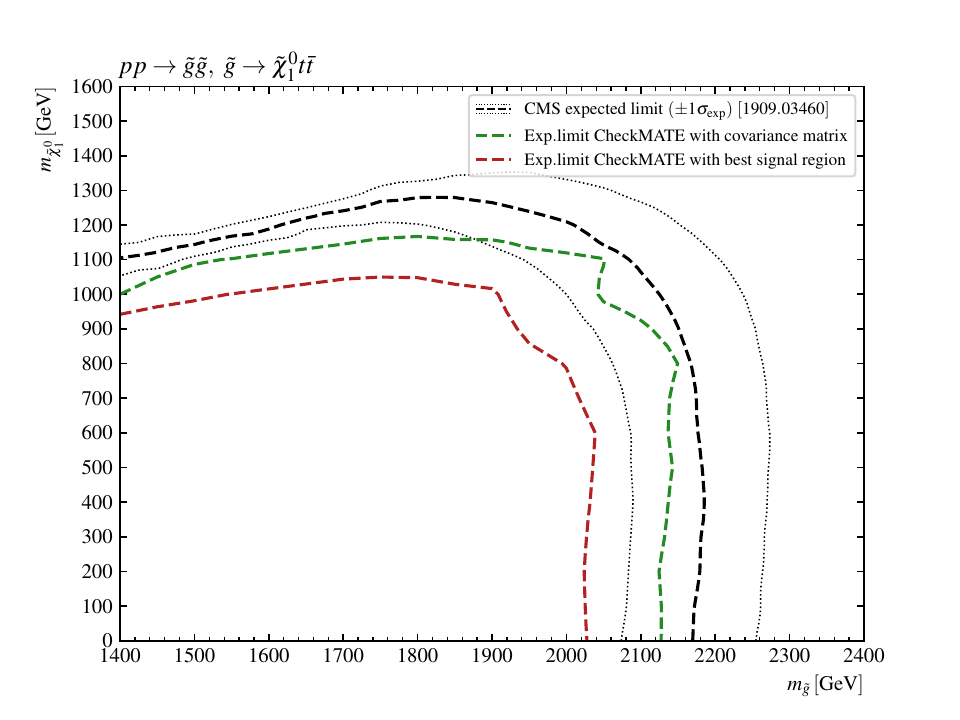}\\
     \caption{Validation plots for the search \texttt{cms\_1909\_03460}. The dotted lines around ATLAS limit denote 1-sigma uncertainty: theoretical for the observed limit and experimental for the expected limit. Left panel: observed limits; right panel: expected limits. \label{fig:190903460}} 
   \end{figure}

\subsection{\texttt{cms\_2107\_13021} (EXO-20-004)}
This is a search \cite{CMS:2021far} for new particles in the final states with at least one jet, no leptons, and missing transverse momentum. A main focus of the analysis are invisible particles that can be dark-matter candidates and that are produced with at least one ISR jet. The simplified likelihood fit is performed on 66 bins: 3 sets for different data-taking periods and divided according to missing transverse energy.

For validation, we compare exclusion limits for the vector mediator model in the $m_\textrm{mediator}$--$m_\textrm{DM}$ plane. The mediator-quark and mediator-DM couplings are fixed to $g_q=0.25$ and $g_\chi = 1$, respectively. The validation plot, Figure~\ref{fig:210713021}, again shows good agreement between the \CM{} and CMS limits.

\begin{figure}
       \includegraphics[width=0.5\textwidth]{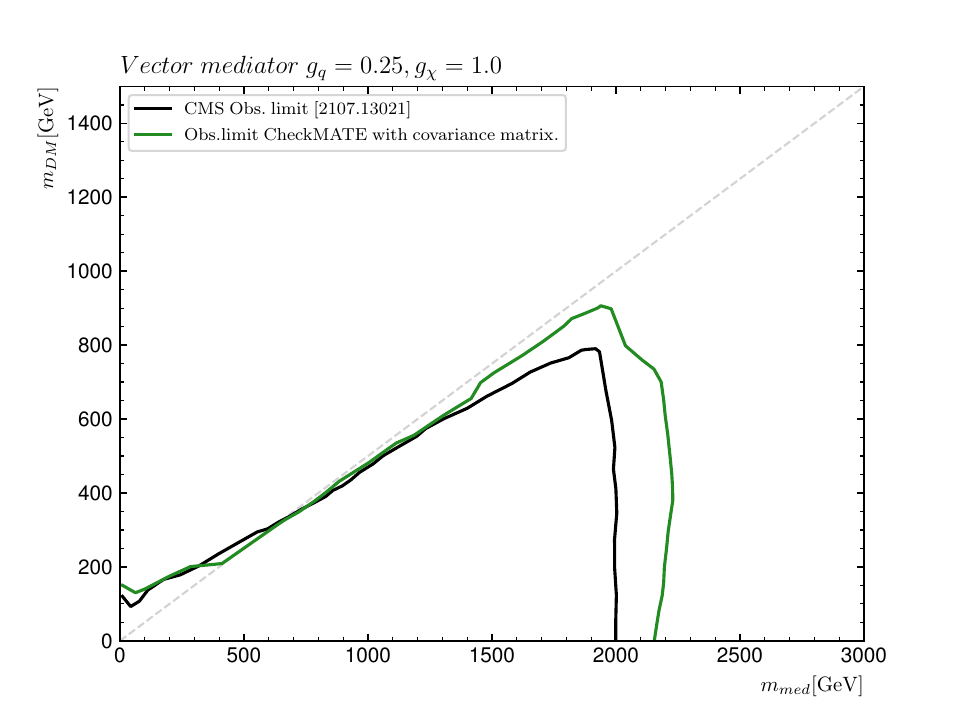}
 \includegraphics[width=0.5\textwidth]{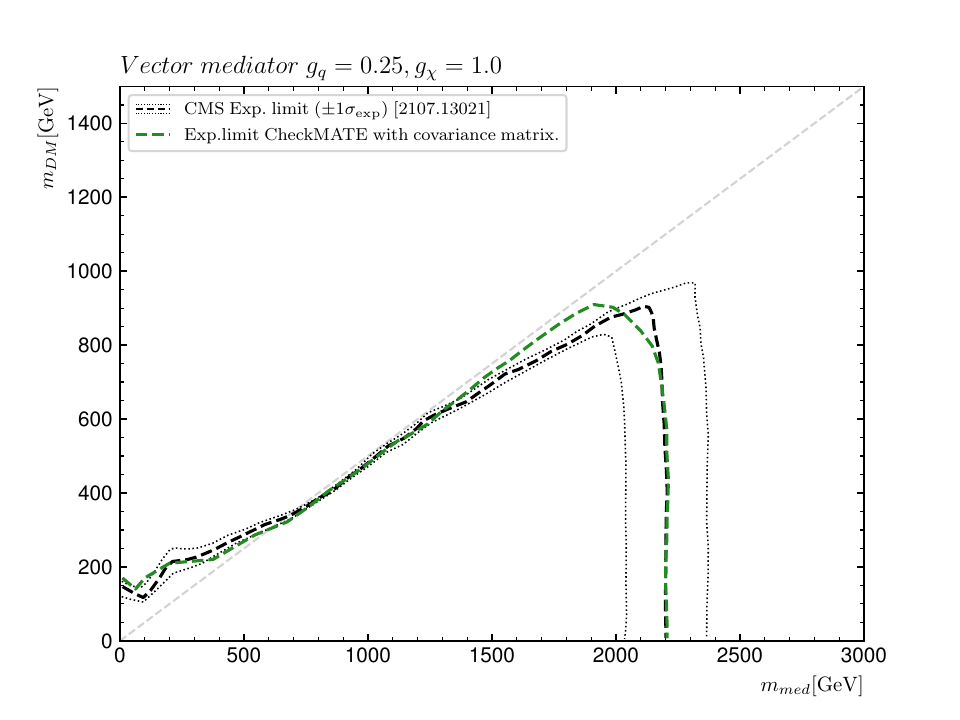}
\caption{Validation plots for the search \texttt{cms\_2107\_13021} and the vector mediator model in the $m_\textrm{mediator}$--$m_\textrm{DM}$ plane. The mediator-quarks and mediator-DM couplings are fixed to $g_q=0.25$ and $g_\chi = 1$, respectively. The diagonal line indicates $m_\textrm{med} = 2 m_\textrm{DM} $.\label{fig:210713021}}
 \end{figure}

\subsection{\texttt{cms\_2205\_09597} (SUS-21-002)}
This is a search \cite{CMS:2022sfi} for charginos and neutralinos that looks for final states with large missing transverse momentum and pairs of hadronically decaying bosons $WW$, $WZ$ and $WH$. This search makes use of specific algorithms (taggers) defined to identify $W$ boson/Higgs boson candidates out of the identified large-radius signal jets. Individual bins are defined orthogonally based on the number of $b$-tagged jets, the number of identified $W$, $Z$, and Higgs boson candidates, and missing transverse momentum. The simplified likelihood is built out of 35 signal regions. Additionally, there are 4 aggregate signal regions defined.

Currently, the implementation is validated for the production of charginos with $W$ bosons in the final state. As can be seen in Fig.~\ref{fig:220509597} there is good agreement between \CM{} and the CMS exclusion contours. The multibin fit clearly improves sensitivity, the red vs.\ green line in the left panel of Fig.~\ref{fig:220509597}. The remaining signal regions for $Z$ and $h$ channels are implemented, however, due to missing mistag rates for different types of final states, they should be used with caution.

 \begin{figure}
        \includegraphics[width=0.5\textwidth]{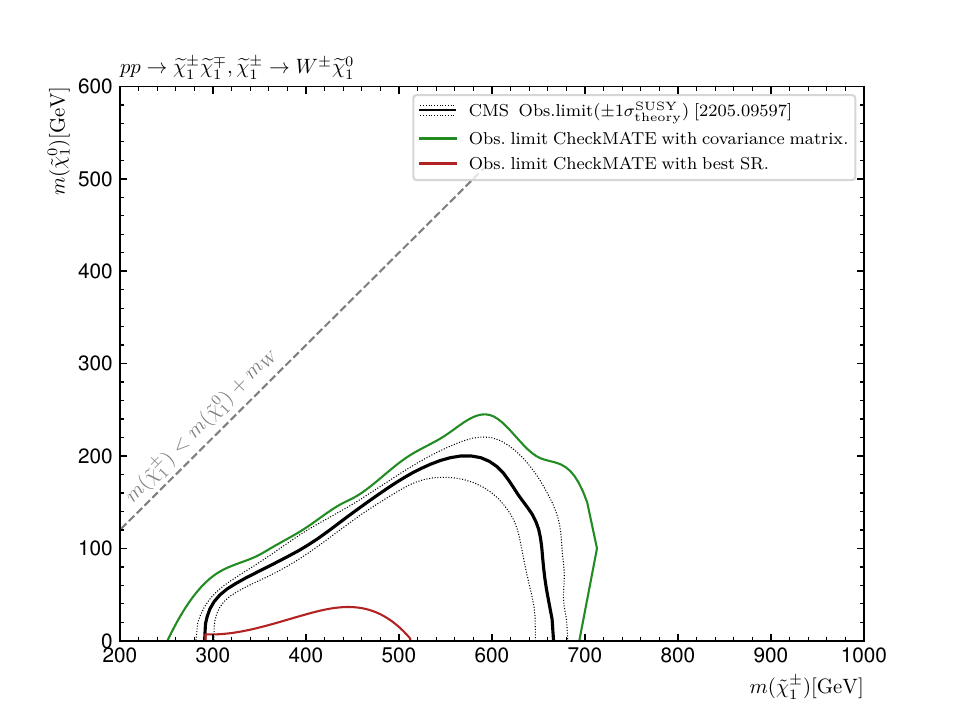}
 \includegraphics[width=0.5\textwidth]{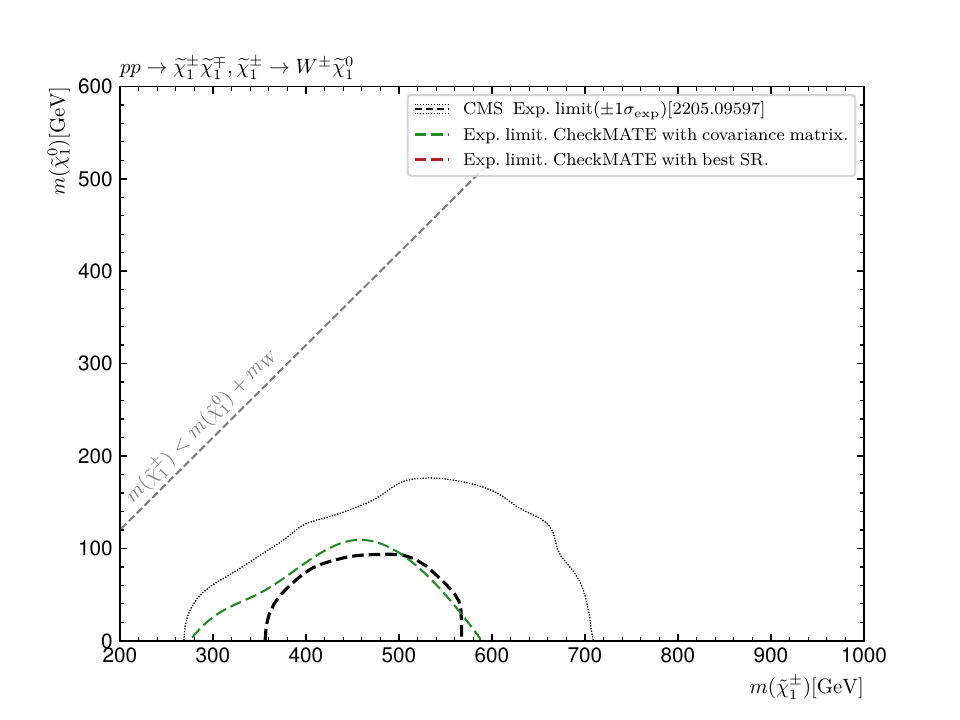}
 \caption{Validation plots for the search \texttt{cms\_2205\_09597} and chargino pair production followed by $\tilde{\chi}^\pm_1 \to W^\pm \tilde{\chi}_1^0$. \label{fig:220509597}}
\end{figure}

\section{Conclusions and outlook} 

In this paper, we present a version of \CM{}, which significantly updates the database of available experimental searches. All included searches are based on the full data set of LHC Run 2. Here, we describe the implementation of 9 ATLAS searches and 4 CMS searches. Another update comes with the inclusion of statistical models, which in a number of examples significantly improve sensitivity and expand exclusion limits. This implementation includes the full and simplified likelihoods provided by the ATLAS Collaboration and the simplified correlated background models provided by CMS. With a number of options available, it allows users to fully control the numerical evaluation. \CM{} can be downloaded from the GitHub repository:
\begin{center}
\url{https://github.com/CheckMATE2/checkmate2/}~~.
\end{center}
More information, including previous versions and expanded validation notes, can be found at:
\begin{center}
\url{http://checkmate.hepforge.org/}~~.
\end{center}

In future releases, we plan to further expand the \CM capabilities by allowing a statistical combination of different searches and orthogonal signal regions. In a separate note, we will cover another new development to \CM{}, which is an interface to machine learning methods which are becoming increasingly common tools in experimental analyses at the LHC.  

\section*{Acknowledgements}
The authors thank Zirui Wang for his help and additional input for Ref.~\cite{ATLAS:2021gcn}. This work was supported by the National Science Centre, Poland under grant 2019/35/B/ST2/02008 and OpenMAPP project under CHIST-ERA programme (grant No.\ NCN 2022/04/Y/ST2/00186). The authors have also received funding from the Norwegian Financial Mechanism 2014-2021, grant 2019/34/H/ST2/00707.

This version of the article has been accepted for publication after peer review, but is
not the Version of Record and does not reflect post-acceptance improvements, or any corrections. The Version of Record is available online at: \url{https://doi.org/10.1140/epjc/s10052-026-15414-8}.

\appendix
\section*{Appendix\label{sec:appendix}}
\subsection*{Event yields for \texttt{atlas\_2010\_14293} (SUSY-2018-22)}
In Tables~\ref{tab:ssd}--\ref{tab:c} we provide a detailed comparison of the event yields in the signal regions of the \texttt{atlas\_2010\_14293} (SUSY-2018-22) search where the MC sample size in \CM{} was 100k. A good agreement (within statistical uncertainties) between ATLAS and \CM{} is observed except for the 4-jet bins in the squark search region \texttt{MB-SSd}. However, the test sample, the squark pair production with direct decays $\tilde{q} \to q \tilde{\chi}_1^0$, at the lowest parton order does not contribute to these bins and requires 2 additional jets from ISR/FSR. It is therefore very sensitive to details in MC modeling and is difficult to reproduce in recasts. These discrepancies do not affect the fit for this signal process due to the small yields compared to expected backgrounds.

\begin{table}\begin{center}\small
 \begin{tabular}{l r r r}
\toprule
Bin & \CM  & ATLAS & ATLAS/\CM \\
\midrule
\texttt{MB-SSd-2-1000-10} & $0.94 \pm 0.033$ & $1  \pm 0.15 $& $1.1 \pm 0.16$ \\
\texttt{MB-SSd-2-1000-16} & $0.74 \pm 0.029$ & $0.59  \pm 0.1 $& $0.8 \pm 0.14$ \\
\texttt{MB-SSd-2-1000-22} & $0.41 \pm 0.022$ & $0.27  \pm 0.071 $& $0.66 \pm 0.17$ \\
\texttt{MB-SSd-2-1600-10} & $3.8 \pm 0.066$ & $3.7  \pm 0.26 $& $0.96 \pm 0.069$ \\
\texttt{MB-SSd-2-1600-16} & $2.2 \pm 0.05$ & $2  \pm 0.18 $& $0.91 \pm 0.085$ \\
\texttt{MB-SSd-2-1600-22} & $4 \pm 0.067$ & $3.5  \pm 0.26 $& $0.88 \pm 0.064$ \\
\texttt{MB-SSd-2-2200-16} & $1.8 \pm 0.045$ & $2  \pm 0.2 $& $1.1 \pm 0.11$ \\
\texttt{MB-SSd-2-2200-22} & $7.8 \pm 0.094$ & $8.4  \pm 0.39 $& $1.1 \pm 0.051$ \\
\texttt{MB-SSd-2-2800-16} & $1.4 \pm 0.04$ & $1.3  \pm 0.16 $& $0.95 \pm 0.11$ \\
\texttt{MB-SSd-2-2800-22} & $5.9 \pm 0.082$ & $6  \pm 0.33 $& $1 \pm 0.055$ \\
\texttt{MB-SSd-2-3400-22} & $1.2 \pm 0.038$ & $0.93  \pm 0.14 $& $0.77 \pm 0.12$ \\
\texttt{MB-SSd-2-3400-28} & $2.6 \pm 0.055$ & $2.2  \pm 0.19 $& $0.84 \pm 0.075$ \\
\texttt{MB-SSd-2-4000-22} & $0.47 \pm 0.024$ & $0.22  \pm 0.063 $& $0.47 \pm 0.13$ \\
\texttt{MB-SSd-2-4000-28} & $0.89 \pm 0.033$ & $0.67  \pm 0.11 $& $0.76 \pm 0.12$ \\
\texttt{MB-SSd-4-1000-10} & $0.13 \pm 0.012$ & $0.63  \pm 0.1 $& $4.7 \pm 0.78$ \\
\texttt{MB-SSd-4-1000-16} & $0.067 \pm 0.0088$ & $0.39  \pm 0.08 $& $5.8 \pm 1.2$ \\
\texttt{MB-SSd-4-1000-22} & $0.015 \pm 0.0041$ & $0.17  \pm 0.064 $& $11 \pm 4.3$ \\
\texttt{MB-SSd-4-1600-10} & $1.7 \pm 0.045$ & $2.4  \pm 0.21 $& $1.4 \pm 0.13$ \\
\texttt{MB-SSd-4-1600-16} & $0.52 \pm 0.025$ & $1.3  \pm 0.16 $& $2.6 \pm 0.3$ \\
\texttt{MB-SSd-4-1600-22} & $0.39 \pm 0.021$ & $1.7  \pm 0.17 $& $4.3 \pm 0.44$ \\
\texttt{MB-SSd-4-2200-16} & $0.77 \pm 0.03$ & $1.4  \pm 0.16 $& $1.8 \pm 0.21$ \\
\texttt{MB-SSd-4-2200-22} & $1.6 \pm 0.042$ & $3.9  \pm 0.26 $& $2.5 \pm 0.17$ \\
\texttt{MB-SSd-4-2800-16} & $1.1 \pm 0.036$ & $0.99  \pm 0.14 $& $0.92 \pm 0.13$ \\
\texttt{MB-SSd-4-2800-22} & $1.9 \pm 0.047$ & $4.6  \pm 0.3 $& $2.4 \pm 0.16$ \\
\bottomrule
\end{tabular}
\caption{Event yields for the squark pair production, $m_{\tilde{q}} = 1700$ GeV, $m_{\tilde{\chi}_1^0} = 500$ GeV with MC statistical uncertainties, $s \pm \Delta s$, normalized to the nominal cross section (8-fold degeneracy) at the approximate NNLO+NNLL accuracy, $\sigma = 0.74$~fb.}
\label{tab:ssd}
 \end{center}
 \end{table}
 
 \begin{table}
 \begin{center}\small
 \begin{tabular}{l r r r}
\toprule
Bin & \CM & ATLAS & ATLAS/\CM \\
\midrule
\texttt{MB-GGd-4-1000-10} & $4 \pm 0.13$ & $2.9  \pm 0.49 $& $0.74 \pm 0.12$ \\
\texttt{MB-GGd-4-1000-16} & $1.4 \pm 0.076$ & $0.64  \pm 0.21 $& $0.46 \pm 0.15$ \\
\texttt{MB-GGd-4-1000-22} & $0.18 \pm 0.027$ & $0.097  \pm 0.097 $& $0.55 \pm 0.55$ \\
\texttt{MB-GGd-4-1600-10} & $17 \pm 0.26$ & $17  \pm 1.3 $& $1 \pm 0.076$ \\
\texttt{MB-GGd-4-1600-16} & $13 \pm 0.23$ & $14  \pm 1.1 $& $1.1 \pm 0.086$ \\
\texttt{MB-GGd-4-1600-22} & $5.1 \pm 0.15$ & $4.4  \pm 0.63 $& $0.85 \pm 0.12$ \\
\texttt{MB-GGd-4-2200-10} & $11 \pm 0.21$ & $13  \pm 1.1 $& $1.2 \pm 0.1$ \\
\texttt{MB-GGd-4-2200-16} & $14 \pm 0.24$ & $17  \pm 1.2 $& $1.2 \pm 0.088$ \\
\texttt{MB-GGd-4-2200-22} & $12 \pm 0.23$ & $14  \pm 1.1 $& $1.1 \pm 0.086$ \\
\texttt{MB-GGd-4-2800-10} & $2.6 \pm 0.11$ & $2.4  \pm 0.42 $& $0.91 \pm 0.16$ \\
\texttt{MB-GGd-4-2800-16} & $3.6 \pm 0.12$ & $3.6  \pm 0.54 $& $1 \pm 0.15$ \\
\texttt{MB-GGd-4-2800-22} & $4.3 \pm 0.13$ & $5.3  \pm 0.65 $& $1.2 \pm 0.15$ \\
\texttt{MB-GGd-4-3400-10} & $0.49 \pm 0.046$ & $0.56  \pm 0.21 $& $1.1 \pm 0.42$ \\
\texttt{MB-GGd-4-3400-16} & $0.47 \pm 0.046$ & $0.3  \pm 0.14 $& $0.64 \pm 0.3$ \\
\texttt{MB-GGd-4-3400-22} & $0.71 \pm 0.056$ & $1.5  \pm 0.36 $& $2.1 \pm 0.51$ \\
\texttt{MB-GGd-4-4000-10} & $0.11 \pm 0.023$ & $0.24  \pm 0.14 $& $2.2 \pm 1.3$ \\
\texttt{MB-GGd-4-4000-16} & $0.11 \pm 0.022$ & $0.11  \pm 0.1 $& $0.98 \pm 0.95$ \\
\texttt{MB-GGd-4-4000-22} & $0.21 \pm 0.031$ & $0.1  \pm 0.1 $& $0.48 \pm 0.48$ \\
\bottomrule
\end{tabular}
\caption{Event yields for the gluino pair production, $m_{\tilde{g}} = 1800$ GeV, $m_{\tilde{\chi}_1^0} = 800$ GeV with MC statistical uncertainties, $s \pm \Delta s$, normalized to the nominal cross section at the approximate NNLO+NNLL accuracy, $\sigma = 2.93$~fb.}

 \end{center}
 \end{table}
 
 \begin{table}
 \begin{center}\small
 \begin{tabular}{l r r r}
\toprule
Bin & \CM  & ATLAS & ATLAS/\CM \\
\midrule
\texttt{MB-C-2-1600-16} & $2.2 \pm 0.05$ & $1.9  \pm 0.18 $& $0.89 \pm 0.084$ \\
\texttt{MB-C-2-1600-22} & $3.9 \pm 0.066$ & $3.3  \pm 0.24 $& $0.85 \pm 0.062$ \\
\texttt{MB-C-2-2200-16} & $2.1 \pm 0.049$ & $2.3  \pm 0.22 $& $1.1 \pm 0.1$ \\
\texttt{MB-C-2-2200-22} & $9.1 \pm 0.1$ & $9.8  \pm 0.42 $& $1.1 \pm 0.047$ \\
\texttt{MB-C-2-2800-16} & $1.7 \pm 0.044$ & $1.5  \pm 0.17 $& $0.91 \pm 0.098$ \\
\texttt{MB-C-2-2800-22} & $9.5 \pm 0.1$ & $9.5  \pm 0.41 $& $1 \pm 0.044$ \\
\texttt{MB-C-4-1600-16} & $0.76 \pm 0.029$ & $0.71  \pm 0.12 $& $0.93 \pm 0.16$ \\
\texttt{MB-C-4-1600-22} & $0.69 \pm 0.028$ & $0.63  \pm 0.11 $& $0.91 \pm 0.16$ \\
\texttt{MB-C-4-2200-16} & $1 \pm 0.034$ & $1.1  \pm 0.14 $& $1.1 \pm 0.14$ \\
\texttt{MB-C-4-2200-22} & $2.9 \pm 0.058$ & $2.5  \pm 0.21 $& $0.86 \pm 0.07$ \\
\texttt{MB-C-4-2800-16} & $1 \pm 0.035$ & $0.76  \pm 0.13 $& $0.73 \pm 0.12$ \\
\texttt{MB-C-4-2800-22} & $4.7 \pm 0.074$ & $4  \pm 0.28 $& $0.84 \pm 0.06$ \\
\texttt{MB-C-5-1600-16} & $0.35 \pm 0.02$ & $0.28  \pm 0.07 $& $0.8 \pm 0.2$ \\
\texttt{MB-C-5-1600-22} & $0.19 \pm 0.015$ & $0.24  \pm 0.068 $& $1.2 \pm 0.35$ \\
\texttt{MB-C-5-2200-16} & $1 \pm 0.034$ & $0.61  \pm 0.11 $& $0.61 \pm 0.11$ \\
\texttt{MB-C-5-2200-22} & $1.6 \pm 0.043$ & $1.2  \pm 0.15 $& $0.77 \pm 0.094$ \\
\texttt{MB-C-5-2800-16} & $1.5 \pm 0.043$ & $0.75  \pm 0.13 $& $0.5 \pm 0.085$ \\
\texttt{MB-C-5-2800-22} & $4.6 \pm 0.074$ & $3.3  \pm 0.25 $& $0.71 \pm 0.055$ \\
\bottomrule
\end{tabular}
\caption{Event yields for the squark pair production, $m_{\tilde{q}} = 1700$ GeV, $m_{\tilde{\chi}_1^0} = 500$ GeV with MC statistical uncertainties, $s \pm \Delta s$, normalized to the nominal cross section (8-fold degeneracy) at the approximate NNLO+NNLL accuracy, $\sigma = 0.74$~fb.}
\label{tab:c}
 \end{center}
 \end{table}

\bibliography{CM_multibin}
\bibliographystyle{utphys28mod}

\end{document}